\documentclass[10pt]{article}

\usepackage{amsmath, amssymb, amsthm, mathtools, xparse, xcolor, amsfonts, soul}
\usepackage{thm-restate}
\usepackage{dsfont}
\usepackage[most]{tcolorbox}
\usepackage[inline]{enumitem}
\usepackage{tikz}
\usepackage{mdframed}
\usepackage[normalem]{ulem}
\usepackage{hyperref}  
\hypersetup{
    colorlinks,
    citecolor=blue,
    filecolor=blue,
    linkcolor=blue,
    urlcolor=blue,
    linktoc=all
}

\usepackage[capitalise]{cleveref}
\crefname{step}{Step}{Steps}
\crefname{condition}{Condition}{Conditions}
\crefname{case}{Case}{Cases}

\usepackage{waingarten}
\newcommand{\ot}{\tilde{O}}

\definecolor{tiansorange}{RGB}{255, 76, 0}

\DeclareFontShape{OT1}{cmr}{m}{scit}{<->ssub*cmr/m/sc}{}

\newcommand{\bo}{\boldsymbol{o}}

\newcommand{\lp}{\left(}
\newcommand{\rp}{\right)}

\newcommand{\minc}{16(6\ln(2/\delta_2)/\delta_1)^{1/p}}

\renewcommand{\close}{\mathsf{CLOSE}}
\renewcommand{\far}{\mathsf{FAR}}

\DeclarePairedDelimiter\paren{(}{)}%
\DeclarePairedDelimiter\brac{[}{]}%
\DeclarePairedDelimiter\ceil{\lceil}{\rceil}%
\DeclarePairedDelimiter\floor{\lfloor}{\rfloor}%

\DeclareMathOperator{\pr}{Pr}

\newcommand{\bAlg}{\boldsymbol{\Alg}}
\newcommand{\pdflp}{\texorpdfstring{$\ell_p$}{lp}}

\newcommand{\br}{\boldsymbol{r}}
\newcommand{\rmi}{\textsc{Random-Multi-Index}}

\newcommand{\AlgCert}{\Alg_{\mathsf{cert}}}

\newcommand{\eventlinkcolor}{darkgray}
\newcommand{\Elink}[1]{
    {\hypersetup{linkcolor=\eventlinkcolor} \hyperref[event:#1]{\calE_{#1}}}
}

\usepackage{breqn}

\title{Average-Distortion Sketching}
\author{
    Yiqiao Bao\thanks{University of Pennsylvania. \texttt{email:~yiqiaob@cis.upenn.edu}
    } \and
    Anubhav Baweja\thanks{University of Pennsylvania. \texttt{email:~abaweja@cis.upenn.edu}
    } \and
    Nicolas Menand\thanks{University of Pennsylvania.
    \texttt{email:~nmenand@cis.upenn.edu}. Supported by the National Science Foundation (NSF) under Grant No. CCF-2337993.
    } \and
    Erik Waingarten\thanks{Univeristy of Pennsylvania.
    \texttt{email:~ewaingar@cis.upenn.edu}. Supported by the National Science Foundation (NSF) under Grant No. CCF-2337993.
    } \and
    Nathan White\thanks{University of Pennsylvania.  
    \texttt{email:~nathanlw@cis.upenn.edu}. Supported by the Department of Defense (DoD) through the National Defense Science and Engineering Graduate (NDSEG) Fellowship Program.} \and
    Tian Zhang\thanks{University of Pennsylvania.
    \texttt{email:~tianzh@cis.upenn.edu}. Supported by the National Science Foundation (NSF) under Grant No. CCF-2337993.
    } 
}
\date{\today}

\begin{document}
\maketitle

\begin{abstract}
We introduce average-distortion sketching for metric spaces. As in (worst-case) sketching, these algorithms compress points in a metric space while approximately recovering pairwise distances. The novelty is studying average-distortion: for any fixed (yet, arbitrary) distribution $\mu$ over the metric, the sketch should not over-estimate distances, and it should (approximately) preserve the average distance with respect to draws from $\mu$. The notion generalizes average-distortion embeddings into $\ell_1$~\cite{R03, KNT21} as well as data-dependent locality-sensitive hashing~\cite{AR15,ANNRW18}, which have been recently studied in the context of nearest neighbor search.
\begin{itemize}
\item For all $p \in (2, \infty)$ and any $c$ larger than a fixed constant, we give an average-distortion sketch for $([\Delta]^d, \ell_p)$ with approximation $c$ and bit-complexity $\poly(2^{p/c} \cdot \log(d\Delta))$, which is provably impossible in (worst-case) sketching.
\item As an application, we improve on the approximation of sublinear-time data structures for nearest neighbor search over $\ell_p$ (for large $p > 2$). The prior best approximation was $O(p)$~\cite{ANNRW18, KNT21}, and we show it can be any $c$ larger than a fixed constant (irrespective of $p$) by using  \smash{$n^{O(p/c)}$} space.
\end{itemize}
We give some evidence that $2^{\Omega(p/c)}$ space may be necessary by giving a lower bound on average-distortion sketches which produce a certain probabilistic certificate of farness (which our sketches crucially rely on). 
\end{abstract}

\newpage

\tableofcontents

\newpage

\section{Introduction}

In this paper, we consider sketching for estimating distances between points in a metric space. For a fixed metric $(X, d_X)$, a sketch is a randomized compression scheme $\sk \colon X \to \{0,1\}^s$ which maps points $x$ of the metric to a small-space sketch of $s$ bits, $\sk(x)$, such that given two sketches $\sk(x)$ and $\sk(y)$, a decoding algorithm can approximately recover $d_X(x,y)$. Historically, sketching metric spaces, and in particular over high-dimensional spaces, was first used for finding near-duplicates and comparing large documents~\cite{B97, BGMZ97, C02}. Over the past decades, sketching has become a fundamental algorithmic tool with broad applications in diverse areas; for example, in streaming algorithms~\cite{M05}, databases~\cite{C11}, nearest neighbor search~\cite{AIR18}, numerical linear algebra~\cite{W14}, and dynamic graph algorithms~\cite{M14}. 

In the context of this work, the most notable case of sketching metric spaces is sketching the $\ell_p$ norms. By now, we have an essentially complete understanding of the entire space-approximation tradeoffs achievable~\cite{BJKS04, IW05}. For every $p \in [0, 2]$, there is a sketch achieving a constant factor approximation (in fact, a $(1+\eps)$-approximation) using $O(\log d/\eps^2)$ many bits; for $p > 2$, the sketching complexity for a $c$-approximation becomes $\tilde{\Theta}(d^{1 - 2/p} / c^2)$, which is polynomial in the underlying dimension for any fixed constant approximation. We are motivated by the following high-level question:
\begin{quote}
Suppose input points are drawn from an arbitrary distribution $\mu$ supported on $(X, d_X)$, do better sketches exist which are tailored to $\mu$?
\end{quote}
The goal of this paper is to define \emph{average-distortion sketching}, a first step at addressing the above question, and to initiate a study of the space-approximation tradeoffs achievable for average-distortion. As we further expand on in Subections~\ref{subsec:avg-dist-sketch} and~\ref{subsec:contrib}, our study will give rise to the following consequences: 
\begin{itemize}
    \item Average-distortion sketching will (strictly) generalize both data-dependent locality-sensitive hashing~\cite{AINR14, AR15, AIR18} and average-distortion metric embeddings to $\ell_1$~\cite{R03, N14}.
    \item Admit sketching algorithms for $\ell_p$ spaces (when $p > 2$) with space-approximation tradeoffs which significantly improve upon what is possible in (worst-case) sketches; worst-case $c$-approximations use $\tilde{\Theta}(d^{1-2/p}/c^2)$ bits, our average-distortion sketches use $\poly(2^{p/c} \log(d))$ bits.
    \item Give rise to new algorithms for approximate nearest neighbor search over $\ell_p$ for $p >2$, which obtain improved approximation factors beyond those which are achievable via (data-dependent) locality-sensitive hashing. 
\end{itemize}
We introduce average-distortion sketching in Definition~\ref{def:avg-sketch}, which roughly speaking, are sketching algorithms which are non-expanding and approximately preserve the average distances among points drawn independently from a distribution $\mu$. Importantly, these guarantees hold without making any assumptions on the distribution $\mu$ itself. For the case of $\ell_p$, a feature of our sketching algorithms is that, even though it is tailored to each distribution $\mu$, the necessary properties can be learned from very few (just $O(\log d)$ many) independent samples from $\mu$. We give some indication that our a space-approximation may be optimal by showing a nearly matching lower bound for sketches which produce a certain type of certificate (which our sketches produce).

\subsection{Related Work}

\paragraph{Data-Dependent Techniques in High-Dimensional Nearest Neighbors.} Traditional algorithms for high-dimensional nearest neighbor search (particularly those based on locality-sensitive hashing) are data-independent. These algorithms proceed by sampling a locality-sensitive hash function and hashing the input dataset and query according to the drawn hash function %
(see the ``Data-Independent Approach'' section in~\cite{AIR18}). Significant progress in locality-sensitive hashing has proceeded by considering an analogous ``distributional'' question: an algorithm considers its input dataset (i.e., the uniform distribution over its dataset) and designs a better locality-sensitive hash family which is tailored to the dataset at hand %
\cite{AINR14, AR15, ALRW17, ARS17, ANNRW18, ANNRW18b, JWZ24}. There is a natural connection between locality-sensitive hashing and sketching, since locality-sensitive hashing gives a specific type of sketch. Given a hash family $\calH$ for $(X, d_X)$, one sketches a point $x \in X$ by drawing $\bh \sim \calH$ and storing a fingerprint of $\bh(x)$. Given two fingerprints of $\bh(x)$ and $\bh(y)$, the algorithm can determine whether a hash collision occurred, i.e., $\bh(x) = \bh(y)$, which is used as a crude estimate of $d_X(x,y)$. In the above reduction, the sketch always uses constant space (since equality protocols are very efficient) but incurs approximation coming from locality-sensitive hashing. A general sketching algorithm is not necessarily restricted in this way; it may use more (yet still sublinear) space to obtain a better approximation.

\paragraph{Average-Distortion Metric Embeddings.} A metric embedding from a metric space $(X, d_X)$ to $(Y, d_Y)$ is a function which maps points in $X$ to points in $Y$ while (approximately) preserving pairwise distances. %
Over the past decades, low-distortion metric embeddings, particularly into $\ell_1$, have emerged as a fundamental tool in the design and analysis of approximation algorithms~\cite{IMS17}. In metric embeddings, the analogous distributional question we study goes under the name \emph{average-distortion embeddings}~\cite{R03, Average-distortion-embedding-into-trees, N14}. %
It turns out, average-distortion can be significantly lower than (worst-case) distortion~\cite{N21}, and from the algorithmic perspective, these better distortions can be leveraged for improved algorithms for nearest neighbor search~\cite{KNT21, AC21}. Similarly to above, there is a natural connection between metric embeddings into $\ell_1$ and sketching, since an embedding into $\ell_1$ gives a specific type of sketch. Given an embedding of $(X, d_X)$ into $\ell_1$ with distortion $\sfD$, one sketches a point $x \in X$ by applying the embedding and then using the constant-size sketch for $\ell_1$~\cite{I06}. The resulting sketch always uses constant size and incurs approximation $O(\sfD)$.\footnote{The sketching via embedding approach is more thoroughly studied in~\cite{AKR15}, who show that constant-size sketches with approximation $\sfD$ might-as-well be $O(\sfD/\eps)$-distortion embeddings into $\ell_{1-\eps}$.} As above, a general sketching algorithm would ideally use more (yet still sublinear) space to achieve better approximations.

\ignore{
\paragraph{Sketching Beyond Worst-Case.} In a parallel line of work, there has been a lot of recent interest designing sketching algorithms which go beyond worst-case. Most related to this work are those on frequency moment estimation, as these correspond to estimating the $\ell_p$-norm of an implicit input vector. Here, improved algorithms have been designed under particular input distributions as well as with predictions on particular properties of the inputs. We propose average-distortion sketches one avenue for sketching 

As we expand on below, our improved sketching algorithm can learn the necessary properties that it needs for improved sketches from few independent samples from $\mu$. Thus, our algorithm falls in the framework of learning-augmented algorithms: for any unknown distribution $\mu$, there exists a learning algorithm which takes a few samples from $\mu$ and produces a sketching algorithm which performs well with respect to $\mu$.}

\subsection{Average-Distortion Sketching}\label{subsec:avg-dist-sketch}
The defining feature of average-distortion sketching is that it assumes knowledge of a fixed, but arbitrary, distribution of points in a metric space. The goal is to compress, i.e., sketch, the elements of the metric while successfully recovering distances (up to some \emph{average-distortion}) with respect to the fixed distribution. Our aim is to design sketching algorithms which use knowledge of the distribution, even though the distribution may be completely arbitrary. Importantly, the definition below is a relaxation of (worst-case) sketching, so we will study average-distortion sketching whenever worst-case sketching is impossible. %

\begin{definition}[Average-Distortion Sketches]\label{def:avg-sketch}
Let $(X, d_X)$ denote a metric space and $\mu$ be a probability distribution supported on $X$. An average-distortion sketch for $\mu$ is specified by a distribution $\calD$ supported on tuples $(\sk, \Alg)$, where $\sk$ is a function and $\Alg$ is an algorithm:
\begin{itemize}
\item The function $\sk \colon X \to \{0,1\}^s$ maps points in the metric space to $s$ bits, and we refer to $s$ as the \emph{space complexity} of the sketch.
\item $\Alg$ is an algorithm which receives as input two sketches $\sk(x)$ and $\sk(y)$ and outputs a number $\boldeta \in \R_{\geq 0}$.
\end{itemize}
We say the sketch achieves average-distortion $c \geq 1$ %
if the following two guarantees are satisfied:
\begin{itemize}
\item \emph{\textbf{Non-Expansion}}: %
For any two points $x,y\in X$,
    $\Ex_{(\sk, \Alg) \sim \calD} [\Alg(\sk(x),\sk(y))] \leq d_X(x,y).$\footnote{In fact, our sketches for $\ell_p$ will have the non-expansion guarantee with high probability. For any $x, y$, $\Alg(\sk(x), \sk(y))$ will be at most $d_X(x,y)$ with probability at least $1-\delta$ over the randomness in $(\sk, \Alg)$ while incurring a $O(\log(1/\delta))$-factor in the space.}%
\item \emph{\textbf{Bounded Contraction}}: We have that
\begin{align*}
\Ex_{\substack{\bx, \by \sim \mu \\ (\sk, \Alg) \sim \calD}}\left[ \Alg(\sk(\bx), \sk(\by)) \right] \geq \frac{1}{c} \cdot \Ex_{\bx, \by \sim \mu}\left[ d_X(\bx, \by) \right],
\end{align*}
where the average is taken over the draw of the sketch as well as independent draws $\bx$ and $\by$ from $\mu$.\footnote{It is important that the expectation is over two independent draws $\bx, \by$ from $\mu$. If the guarantee was over arbitrary an distribution $\mu$ over pairs of vectors $(\bx,\by)$ which are not necessarily independent, we would obtain by duality a worst-case sketch and run into the lower bounds we seek to overcome.}
\end{itemize}
\end{definition}

In what follows, we discuss aspects of Definition~\ref{def:avg-sketch} and draw a comparison to average-distortion embeddings, first introduced in~\cite{R03} (see also \cite{N14,KNT21}). In particular, given a metric space $(X,d_X)$ and a distribution $\mu$ supported on $X$, an average-distortion embedding $f\colon X \to \ell_1$ for $\mu$ into $\ell_1$ is a function satisfying the following two analogous guarantees:
\begin{itemize}
    \item The function is non-expanding, in the sense that $\|f(x) - f(y)\|_1 \leq d_X(x,y)$ for all $x, y \in X$.  This condition is equivalent to requiring $\Ex_{\boldf}\left[ \|\boldf(x) - \boldf(x)\|_1 \right] \leq d_{X}(x,y)$ for random functions $\boldf \colon X \to \ell_1$ by concatenating coordinates.
    \item The average-contraction over $\mu$ is bounded, in the sense that the expectation of $\| f(\bx) - f(\by)\|_1$ where $\bx$ and $\by$ are drawn independently from $\mu$ is at least $1/c$ times the expectation of $d_X(\bx,\by)$, where $\bx,\by\sim \mu$. 
\end{itemize}
Average-distortion sketches in Definition~\ref{def:avg-sketch} generalize average-distortion embeddings in the same way that sketching generalizes embeddings. Unlike embeddings, the function $\sk$ in the sketch is not constrained to mapping points of $X$ to $\ell_1$ (or even a metric); it is allowed to use an arbitrary encoding of the points, so long as pairs of sketches can be decoded. In particular, one approach to design average-distortion sketches is to use an average-distortion embedding into $\ell_1$, and then use a (worst-case) sketch for $\ell_1$. 

\paragraph{Non-Expansion for All Pairs.} Average-distortion sketching imposes a worst-case guarantee on the expected expansion of any pair of points.\footnote{Our upper bounds will also be non-expanding with probability $1-\delta$ for any pair of points, up to $O(\log(1/\delta))$-factor in space.} This guarantee is analogous to non-expansion in average-distortion embeddings, and is crucial for our application to nearest neighbor search. Roughly speaking, in approximate nearest neighbor search, an unknown query $q$ is promised to be nearby a dataset point $x$. Even though the entire dataset is known during preprocessing (so the distribution $\mu$ used will be uniform over the dataset), the query $q$ is arbitrary and unknown to the preprocessing algorithm, and thus, the true nearest neighbor $x$ in the dataset will also be a (arbitrary and non-random) dataset point. At a high level, we want the sketch $(\sk, \Alg)$ to still interpret $\sk(q)$ and $\sk(x)$ as being ``close'' to each other even though the pair $(q, x)$ may be arbitrary. Ensuring non-expansion for worst-case pairs of points $x,y$ allows us to argue this fact.

\paragraph{Worst-Case over Distributions.} Even though average-distortion sketching is tailored to a particular distribution $\mu$, we say that a metric $(X, d_X)$ admits an average-distortion sketch of space-approximation tradeoff $s$ versus $c$ only if an average-distortion sketch exists for \emph{all} distributions $\mu$ over $(X, d_X)$. Thus, the notion of average-distortion sketching is still a worst-case notion. The subtlety is the following: the guarantees on the contraction are not worst-case over pairs of points, since there may be pairs $x, y \in X$ where $\Alg(\sk(x), \sk(y))$ is much smaller than $d_X(x, y) / c$; however, the average-distortion is worst-case over distributions $\mu$. This makes average-distortion sketching applicable to (worst-case) approximate nearest neighbor search data structures. %
Definition~\ref{def:avg-sketch} also suggests a (stronger) model of probabilistic sketching, which we leave for future work: beyond maintaining the non-expansion property, the sketch additionally requires a stronger bounded contraction guarantee: with probability approaching 1 over the inputs and the draw of the sketch, $\Alg(\sk(\bx), \sk(\by)) \geq d_X(\bx,\by)/c$ for independent draws $\bx,\by \sim \mu$.

\subsection{Our Contributions}\label{subsec:contrib}

\paragraph{Average-Distortion Sketch.} Our main result is an average-distortion sketching algorithm for $\ell_p$ for $p > 2$. Importantly, our space-approximation tradeoffs are provably impossible for both (i) sketching $\ell_p$ with $p > 2$ in the worst-case, as well as (ii) average-distortion embeddings from $\ell_p$ to $\ell_1$. Thus, average-distortion sketching strictly generalizes these two notions and admits significantly better algorithms. As we will show, the relaxed guarantee of average-distortion sketching compared to worst-case sketching gives improved algorithms for approximate nearest neighbor search under $(\R^d, \ell_p)$ when $p > 2$.

\begin{theorem}[Average-Distortion Sketching for $\ell_p$]\label{thm:ell-p-sketch}
    For any $c$ greater than a fixed universal constant and any $p \in (2, \infty)$, there exists an average-distortion sketch for any distribution over $([\Delta]^d, \ell_p)$ with distortion $c$ %
    using $2^{O(p/c)}\cdot\log^2(d\Delta) $ bits.\footnote{In fact, for any $x,y \in [\Delta]^d$, a sketch using $2^{O(p/c)}\log^2(d\Delta/\delta)$ bits will satisfy $\Alg(\sk(x), \sk(y)) \leq d_X(x,y)$ with probability $1-\delta$.}
    
\end{theorem}

As a point of comparison, recall that the space complexity of sketching $\ell_p^d$ with $p > 2$ in (worst-case) sketching is \smash{$\tilde{\Theta}(d^{1 - 2/p} / c^2)$} for a $c$-approximation. Hence, optimizing for the average-distortion sketching enables significant space savings. In particular, instantiating Theorem~\ref{thm:ell-p-sketch} for any constant $p > 2$ gives a sketch using poly-logarithmic bits for a constant approximation, whereas constant-approximation (worst-case) sketches require space polynomial in $d$. The other related point of comparison are average-distortion embeddings of $\ell_p$ into $\ell_1$, where the distortion achievable is $\Theta(p)$~\cite{M97, KNT21}, and should be considered as implying average-distortion sketching using $O(\log(d\Delta))$ bits (by sketching $\ell_1$). On the other hand, Theorem~\ref{thm:ell-p-sketch} can achieve average-distortion down to constant $c$ (irrespective of $p$) while using more (yet still sublinear) space. 

\paragraph{Application to Nearest Neighbor Search.} We show how to use average-distortion sketching to design algorithms for nearest neighbor search over $\ell_p$ with better approximation factors.
In particular, the proof of Theorem~\ref{thm:ell-p-sketch} proceeds by building a ``single-scale'' average-distortion sketch with bit-complexity which is completely independent of $d$ and $\Delta$ (Lemma~\ref{lem:boost-prob-of-non-expansion}), which can distinguish between pairs of points within distance $r$, and draws from the distribution at distance at least $cr$.
Average-distortion sketches do not directly imply data-dependent locality-sensitive hash families, yet we show these are still applicable in nearest neighbor search. %
Furthermore, these sketches can be made ``asymmetric'' with minimal modifications: while the sketch of one point uses $2^{\Theta(p/c)}$ space, the other point can be sketched with space only $O(p/c)$ (\cref{corr:asymmetric-sketch}), which allows us to further reduce the space used in the ANN data structure.

\begin{theorem}[Approximate Nearest Neighbor in $\ell_p$]\label{thm:nns-p}
    For any $c$ greater than a fixed universal constant and any $p \in (2, \infty)$ and $\eps > 0$, there is a data structure for $c$-approximate nearest neighbor over $\ell_p$ with the following guarantees:
    \begin{itemize}
        \item \emph{\textbf{Query Time}}: The time to execute a single query is $n^{\eps}d \cdot \poly(\log nd)$.
        \item \emph{\textbf{Space and Preprocessing Time}}: The preprocessing time and space is $d\cdot n^{O(p/c)\cdot \log(1/\varepsilon)}$.
    \end{itemize}
\end{theorem}

Theorem~\ref{thm:nns-p} gives the best approximation factor for nearest neighbor search in $\ell_p$ spaces by using more space in the data structure. Prior to this work, nearest neighbor search for $\ell_p$ had an approximation factor of $O(p/\eps)$ for query time $\poly(d) \cdot n^{\eps}$ and space $\poly(d) \cdot n^{1 + \eps}$~\cite{ANNRW18, KNT21}, improving on a $2^{O(p)}$ approximation of~\cite{BG19}. For large constant $p$, Theorem~\ref{thm:nns-p} can give a constant-factor approximation which is independent of $p$ while searching in sublinear time and using polynomial space (by setting $\eps$ to be a small constant). 

The other point of comparison are the decomposition techniques in~\cite{I01,ANRW21}. For example,~\cite{I01} gives a $O\lp\log_{\rho}\log d\rp$-approximation data structure for $\ell_{\infty}^d$ in sublinear time and $n^{\rho}$ space, which can be made a constant-approximation with \smash{$n^{O(\log d)}$} space by setting $\rho = \log d$. Applying a randomized embedding of $\ell_p^d$ to $\ell_{\infty}^{d}$ (see~\cite{ANNRW17}), it can give a constant-factor approximation using $n^{O(\log d)}$ space. For constant $p$, Theorem~\ref{thm:nns-p} shows this same result can be accomplished in polynomial space, as opposed to \smash{$n^{O(\log d)}$} space.
{Simultaneously, for $\ell_\infty$, Theorem \ref{thm:nns-p} matches \cite{I01} asymptotically in this regime, by setting $p=\log d$ and $c=O(1)$ (since $\ell_{\log d}$ approximates $\ell_\infty$ to within a constant factor).}

\paragraph{Optimality of Theorem~\ref{thm:ell-p-sketch}.} Currently, we do not have strong lower bounds which rule out better space-approximation tradeoffs for average-distortion sketching of $\ell_p$ beyond that which is achievable by Theorem~\ref{thm:ell-p-sketch}. However, the exponential dependence on $p/c$ is intrinsic to our approach. In particular, the ``single-scale'' version of Theorem~\ref{thm:ell-p-sketch} (distinguishing distance $r$ versus $cr$ for independent draws of $\mu$) proceeds via the following argument. It shows how, for any distribution $\mu$ where $\bx, \by \sim \mu$ are likely to be at distance at least $cr$, given $\sk(x)$ and $\sk(y)$, it can produce with constant probability over $\bx,\by\sim \mu$ a certain ``probabilistic certificate'' for the assertion $\|\bx - \by\|_p > r$; furthermore, had $x$ and $y$ been (worst-case) input points at distance at most $r$, the probability that such a certificate would exist is at most $\delta$ (over the draw of $(\sk, \Alg)$). In Section~\ref{sec:lb}, we give an example distribution $\mu$ {supported on integer vectors} where $\bx, \by \sim \mu$ have distance at least $c$, and where our protocol is able to identify, with constant probability, a coordinate $\bi \in [d]$ and an integer threshold $\btau$ where $\bx_{\bi} < \btau < \by_{\bi}$; note that such inequality (for any $x, y$) would certify that $\| x - y\|_p > 2$. We prove that any sketch which can recover such certificates $(\bi,\btau)$ must use $2^{\Omega(p/c)}$ bits (Theorem~\ref{thm:lb}).

\newcommand{\ED}{\mathsf{ED}}
\paragraph{Open Problems.} We believe our relaxation, average-distortion sketching, is a fruitful avenue for overcoming long-standing barriers in the complexity of sketching metric spaces, and their corresponding algorithmic applications.
Toward that end, we ask the following open problems:
\begin{itemize}
    \item \textbf{EMD}. For $s,\Delta \in \N$, let $\EMD_s([\Delta]^d, \ell_1)$ denote the metric space over size-$s$ subsets of $[\Delta]^d$ where the distance between two size-$s$ subsets $x, y \subset [\Delta]^d$ is the Earth Mover's Distance over $\ell_1$~\cite{CJLW21, JWZ24}. Does there exists an average-distortion sketch with distortion $o_{\eps}(\log s)$ and space $s^{\eps} \cdot \poly(d \log(\Delta))$? Currently, (worst-case) sketches achieve $O(\log^2 s)$-approximation using $\polylog(s\Delta d)$ space, and~\cite{JWZ24} implies a $\tilde{O}(\log s)$-average distortion sketch with $\polylog(sd\Delta)$ space. Does there exists an average distortion sketch for $\EMD$ over $[\Delta]^2$ with $\polylog(\Delta)$ space and constant approximation? Note~\cite{ABIW09} give (worst-case) sketches with $O(1/\eps)$-factor approximations and $\Delta^{O(\eps)}$ space. 
    \item \textbf{Edit Distance.} The metric $\ED(d)$ is defined over strings in $\{0,1\}^d$, where $\ED(x, y)$ is the minimum number of insertions, deletions, and substitutions needed to turn $x$ into $y$. The best (worst-case) sketch uses $\polylog(d)$-space and incurs approximation $2^{\tilde{O}(\sqrt{\log d})}$~\cite{OR07,AO09}. Do there exist average-distortion sketches with smaller approximation yet sublinear space? Note, the best sublinear time algorithm for nearest neighbor search over $\ED(d)$ incurs approximation $2^{\tilde{O}(\sqrt{\log d})}$, and our hope is average-distortion sketches can be an approach toward obtaining better approximations.
\end{itemize}

\subsection{Technical Overview}

We give an informal overview of the average-distortion sketch for $\ell_p$ of Section~\ref{sec:ell-p}. For simplicity in the exposition, we describe the sketch for the ``single-scale'' decision version of the problem, i.e., deciding between pairs of points at distance at most $1$ versus independent draws at distance at least $c$ for $([\Delta]^d, \ell_{\infty})$ (Lemma~\ref{lem:single-scale}). Note that $\ell_{\infty}$ in a $d$-dimensional space is up to a constant factor $\ell_p$ with $p = O(\log d)$, which means that Theorem~\ref{thm:ell-p-sketch} would give a sketch of size $d^{O(1/c)}$.

Perhaps surprisingly, the sketch uses minimal knowledge of the (arbitrary) distribution $\mu$ over $([\Delta]^d, \ell_{\infty})$. We only require knowledge of the coordinate-wise median of the distribution. After a translation by the median, we can ensure that, for any coordinate $i \in [d]$, the probability that $\by_i \leq 0$ is at least $1/2$ over $\by \sim \mu$.
Assume this is the case, and consider any fixed vector $x \in \{-\Delta, \dots, \Delta\}^d$ which satisfies $\|x\|_{\infty} \geq c/2$, and furthermore, that the maximizing coordinate $j$ satisfies $x_{j} \geq c/2 > 0$ is positive.\footnote{The distribution is now supported on $\{-\Delta, \dots, \Delta\}^d$ after the translation, and one must formally consider the argument for both possible signs of $x_j$.} We design a sketching algorithm such that, from $\sk(x)$ and $\sk(\by)$ for $\by \sim \mu$, it tries to find a ``certificate of farness'', i.e., a coordinate $\bi \in [d]$ and a threshold $\btau \in [\Delta]$ such that $x_{\bi} \geq \btau$ and $\by_{\bi} < \btau - 1$. If such a certificate is found, the sketch will (safely) output $\far$, since it implies $\|x - \by\|_{\infty} \geq |x_{\bi} - \by_{\bi}| > 1$. Otherwise, it outputs $\close$.

\textbf{Histogram of Coordinates}. For $\sfL = c/2$, we denote the sets of coordinates $G_1(x), G_2(x), \dots, G_{\sfL}(x)$ where the set $G_{\ell}(x)$ consists of coordinates $i \in [d]$ where $x_i \geq \ell$. Notice that these sets are nested, so $G_{\sfL}(x) \subset \dots \subset G_{1}(x)$, and that $1 \leq |G_{\sfL}(x)|$ (by assumption that $x_{j} \geq c/2$) and $|G_1(x)| \leq d$. Roughly speaking, for each coordinate $i \in G_1(x)$, with $1/2$ probability over $\by \sim \mu$, we have $\by_i \leq 0$ 
and thus $i\not\in G_1(y)$. If the sketch $\sk(x)$ manages to identify a coordinate $i^* \in G_{\ell}(x)$ such that $\sk(\by)$ also contains a ``proof'' that $i^{*} \notin G_{\ell-1}(y)$, then we obtain the desired certificate. The challenge is that $\sk(x)$ and $\sk(\by)$ cannot directly cooperate in storing the same coordinate $i^*$.

\textbf{When $G_{\ell-1}(\by)$ is not much larger than $G_{\ell}(x)$}. Suppose that $\sk(x)$ picks a coordinate $\bi_{\mathsf{xmin}} \in G_{\ell}(x)$, and stores that coordinate $\bi_{\mathsf{xmin}} \in [d]$ and the threshold $\ell \in [\sfL]$. A key observation is that a ``proof'' that $\bi_{\mathsf{xmin}} \notin G_{\ell-1}(\by)$ does not necessarily need to store $\bi_{\mathsf{xmin}}$. First, consider the case that $G_{\ell-1}(\by)$ is not too much larger than $G_{\ell}(x)$; namely, for a parameter $\sfk > 1$ (which we set later), 
\begin{align} |G_{\ell-1}(\by)| \leq (\sfk/4) \cdot |G_{\ell}(x)| \enspace. \label{eq:size-compare}
\end{align}
Let $\bpi \colon [d] \to [d]$ be a random permutation, and let $\bi_{\mathsf{xmin}}$ be the first coordinate of $G_{\ell}(x)$ with respect to the random permutation $\bpi$. 

Suppose that $\sk(\by)$ stores the first $\sfk$ coordinates of $G_{\ell-1}(\by)$ with respect to $\bpi$, and that these $\sfk$ coordinates do not contain $\bi_{\mathsf{xmin}}$ (which should happen, roughly speaking, with probability $1/2$ over $\by\sim\mu$). Then, an algorithm $\Alg(\sk(x), \sk(\by))$ which knows that (\ref{eq:size-compare}) is satisfied can confidently assert that $x_{\bi_{\mathsf{xmin}}} \geq \ell$ and that $\by_{\bi_{\mathsf{xmin}}} < \ell - 1$.

From the algorithm's perspective, had $\bi_{\mathsf{xmin}}$ been inside $G_{\ell-1}(\by)$, the probability (under the random permutation $\bpi$) that $\bi_{\mathsf{xmin}}$ was not among the first $\sfk$ elements of $G_{\ell-1}(\by)$ is exponentially small in $\sfk$. 
This is because this event only occurs when the number of other coordinates $i \in G_{\ell-1}(\by)$ which satisfy $\bpi(i) < \bpi(\bi_{\mathsf{xmin}})$ is at least $\sfk$.
The number of such coordinates is at most $\sfk/4$ in expectation (since condition (\ref{eq:size-compare}) holds), and is tightly concentrated since $\bpi$ is chosen uniformly at random.
Hence, if $\sk(x)$ and $\sk(\by)$ jointly sample $\bpi$ and store the first $\sfk$ coordinates in $G_{\ell}(x)$ and $G_{\ell}(\by)$ with respect to $\bpi$ for all $\ell \in [\sfL]$, as well as the sizes of $|G_{\ell}(x)|$ and $|G_{\ell}(\by)|$, the algorithm can check condition (\ref{eq:size-compare}) and output $\far$ confidently.

\textbf{When $G_{\ell-1}(\by)$ is strictly larger than $G_{\ell-2}(x)$}. In the above step, the algorithm crucially relied on the condition (\ref{eq:size-compare}) being satisfied, which may not occur.
In this case, $|G_{\ell-1}(\by)| > (\sfk / 4) \cdot |G_{\ell}(x)|$, but now suppose that the following inequality holds:
\begin{align}
    |G_{\ell-1}(\by)| > |G_{\ell-2}(x)| \enspace. \label{eq:size-compare-2}
\end{align}
So, there exists a coordinate $i \in [d]$ such that $i \in G_{\ell-1}(\by) \setminus G_{\ell-2}(x)$.
Therefore, $x_i < \ell - 2$ but $\by_i \geq \ell - 1$, and we have $\|x - \by\|_{\infty} \geq |x_i - \by_i| > 1$.
Therefore, $\Alg$ can output $\far$.

\textbf{Neither (\ref{eq:size-compare}) nor (\ref{eq:size-compare-2}) hold}. In this case, we can combine the inequalities of failure of (\ref{eq:size-compare}) and (\ref{eq:size-compare-2}) to say
\[ |G_{\ell-2}(x)| \geq |G_{\ell-1}(\by)| > (\sfk/4) \cdot |G_{\ell}(x)|. \]
Recall that $G_{\sfL}(x)$ has size at least $1$, so the failure of (\ref{eq:size-compare}) and (\ref{eq:size-compare-2}) with $\ell = \sfL$ implies $|G_{\sfL - 2}(x)| \geq \sfk / 4$.
Similarly, if conditions (\ref{eq:size-compare}) and (\ref{eq:size-compare-2}) fail again with $\ell = \sfL-2$, we have $|G_{\sfL - 4}(x)| \geq (\sfk/4)^{2}$.
In general, failure of at least, say, $\sfL/10$ values in $[\sfL]$ implies that $|G_{1}(x)| \geq (\sfk / 4)^{\sfL / 20}$.
Recall that $\sfL = \Theta(c)$, so if we set $\sfk = d^{\Theta(1/c)}$, one would conclude that $G_1(x)$ is larger than $d$.
This is a contradiction, and implies that, for many $\ell \in [\sfL]$ one of the two conditions (\ref{eq:size-compare}) or (\ref{eq:size-compare-2}) must hold.
Formally, some care must be taken in the argument, so that the above ``case analysis'' is not inadvertently conditioning on a draw of $\by \sim \mu$ (proof of \cref{lem:prob-close-small}). This is important since the case of (\ref{eq:size-compare}) requires $\by_{\bi_{\mathsf{xmin}}}$ and $x_{\bi_{\mathsf{xmin}}}$ to have opposite signs, which happens with probability $1/2$ (as long as we do not condition on any event). 

\newcommand{\Exp}{\mathrm{Exp}}

\textbf{Extension to the asymmetric setting.}
Note that in the above algorithm, the only information that $\Alg$ needs about $x$ is the coordinate $\bi_{\mathsf{xmin}}$, its $\ell_p$ norm, and the sizes of the sets $G_{\ell-2}(x)$ and $G_{\ell}(x)$.
This is particularly useful when we discuss the ``asymmetric sketch'' in \cref{sec:asymmetric-sketch}, which is useful for our application to the approximate nearest neighbor search problem \cref{sec:applications}.

\textbf{Extension to $\ell_p$.}
In order to extend the above argument to $\ell_p^d$, we apply a probabilistic embedding from \smash{$\ell_p^d \to \ell_{\infty}^d$}, which sets \smash{$\bx_i' = x_i / \bu_i^{1/p}$} for $\bu_i \sim \Exp(1)$. This embedding preserves zero as the coordinate-wise median of the distribution and plays well with our analysis for the following reason:
\begin{itemize}
    \item First, any two $x, y$ with $\|x-y\|_p \leq 1$ will have every coordinate $|\bx_i' - \by_i'| \leq 1 / \delta_1^{1/p}$ except with probability $\delta_1$ (Lemma~\ref{lem:close}). In the above technical overview, we considered thresholds $\ell \in [\sfL]$ so two neighboring thresholds differ by at least 1. Now, we must change thresholds $\ell$ so that any two neighboring thresholds differ by at least \smash{$1 / \delta_1^{1/p}$}.
    \item Second, if $\|x\|_p \geq c$, there will be at least one coordinate where $\bx_i' \geq \|x\|_p / \sfD_1$, and at most $2^{O(p)}$ coordinates where $\bx_i' \geq \|x\|_p / \sfD_2$, for constant $\sfD_1$ and $\sfD_2 = 2 \sfD_1$ with high probability (Lemma~\ref{lem:good-event}). Hence, the contradiction arises by setting the largest threshold $\ell$ to $\|x\|_p / \sfD_1$ (which was $\sfL$ above) and moving toward the threshold $\|x\|_p / \sfD_2$ which contains at most $2^{O(p)}$ coordinates (instead of at most $d$ coordinates).
\end{itemize}
A minor complication which arises, however, is that $\sk(x)$ and $\sk(\by)$ must coordinate on which thresholds to consider. In the overview for $\ell_{\infty}$, the thresholds were $1, 2, \dots, c/2$, which after the embedding, should start at $\|x\|_p/\sfD_2$ and end at $\|x\|_p / \sfD_1$ while differing by \smash{$1/\delta_1^{1/p}$}. These depend on $\|x\|_p$ which $\sk(\by)$ does not know. The plan is for sketches $\sk(x)$ and $\sk(\by)$ to round the norms of $\|x\|_p$ and $\|\by\|_p$ (both up and down) to the nearest multiple of \smash{$1/\delta_1^{1/p}$}---call these $\nu_x$ and $\nu_{\by}$---and begin the thresholds at $\nu_x / \sfD_2$ and $\nu_{\by} / \sfD_2$. If $\left|\|x\|_p - \|y\|_p \right| \leq 1$, at least one of the rounded norms agrees and thresholds align. If, however, the rounded norms never agree, then it means $1 < \left|\|x\|_p - \|y\|_p \right| \leq \|x - y\|_p$, and is another ``proof'' where the sketch may safely output $\far$.

\textbf{Eliminating $\log d$ factors in space.}
In this overview, we mentioned that the sketch stores indices of coordinates (such as $\bi_{\mathsf{xmin}}$), and also stores sizes of sets of coordinates (such as $|G_{\ell}(x)|, |G_{\ell-1}(\by)|, |G_{\ell-2}(x)|$).
These values require $O(\log d)$ bits to store, and therefore pose a roadblock to achieving the single-scale space complexity of $2^{O(p/c)}$ bits (which is independent of $d$).
However, $\Alg$ does not need to know the exact values of these indices or sizes.
In particular, our full construction hashes the indices to a universe of size $\Theta(\sfk) = 2^{\Theta(p/c)}$, and skips the storage of the sizes of sets.
This only leads to a small additional error probability in the case where $x$ and $\by$ have distance more than $c$.

\ignore{

Our primary motivation for Definition~\ref{def:avg-sketch} is generalizing the notion of average-distortion embeddings into $\ell_1$ in order to obtain better approximations using more (yet still sublinear) space. Such average-distortion embeddings have applications in desining data-dependent locality-sensitive hashing and thus approximate nearest neighbor data structures~\cite{AR15, KNT21, AC21}, and our goal is to derive improved approximation factors by considering average-distortion sketching. As discussed, (worst-case) sketches for the $\ell_p$ have space complexity $\tilde{\Theta}(d^{1-2/p} / c^2)$ for approximation $c > 1$. The existence of average-distortion embeddings from $\ell_p \to \ell_1$ is coupled with Indyk's sketch for $\ell_1$~\cite{I06} gives the following corollary. 
\begin{corollary}[Known from~\cite{KNT21}]\label{corr:l-p}
For $p \in [1, \infty)$, there exists an average-distortion sketch for any $\mu$ over $([\Delta]^d, \ell_p)$ with distortion $O(p)$ and failure probability $\delta$ which has space complexity $O(\log(d\Delta/\delta))$.
\end{corollary}

From (worst-case) sketches for $\ell_1$, average-distortion embeddings into $\ell_1$ imply dramatic compressions (i.e., from $d \log \Delta$ to $O(\log(d\Delta))$ bits in Corollary~\ref{corr:l-p}). In this light, Theorem~\ref{thm:avg-sk-l-p} is an example improved approximations given by larger, yet still sublinear space complexity, ``sketch spaces''. Furthermore, distortions beyond $p$ of the type in Theorem~\ref{thm:avg-sk-l-p} cannot be achieved via a better average-distortion embeddings into $\ell_1$. In particular, any average-distortion embedding from $\ell_p$ to $\ell_1$ must incur distortion $\Omega(p)$. This follows from Theorem~1 in~\cite{M97}, all $n$-point metrics embed into $\ell_p$ with distortion $O(\log n / p)$, and there are $n$-point metrics (namely, shortest path metrics on constant-degree expanders) which incur average-distortion $\Omega(\log n)$ into $\ell_1$~\cite{LLR95,M97}.

As we show in Section~\ref{sec:ell-p}, the achievable distortion on average-distortion sketching can be significantly improved by allowing more space. If, instead of requiring compressions of $d$-dimensional vectors to $O(\log(d\Delta))$ bits, we allow ourselves slightly more space, the average-distortion decreases. We show that Corollary~\ref{corr:l-p} is merely one point within achievable algorithms which tradeoff compression for average distortion. The following is our main algorithmic theorem.
\begin{theorem}\label{thm:avg-sk-l-p}
For $p \in [1, \infty)$ and any $c \geq \minc$, there exists an average-distortion sketch for $\mu$ over $([\Delta]^d,\ell_p)$ with distortion $c$ and failure probability $\delta$ which has space complexity $\poly(c \cdot p \cdot 2^{p/c}) \cdot \log(d\Delta/\delta)$.
\end{theorem}

We conclude the section with a few remarks comparing the average-distortion sketching model to the regular (worst-case) sketching model, and a more thorough comparison between average-distortion embedding and average-distortion sketching.

\paragraph{Average-Distortion vs (Worst-Case) Sketching.} The formal definition of (worst-case) sketching is similar to Definition~\ref{def:avg-sketch}, without an underlying distribution $\mu$. A sketch is a distribution $\calD$ supported on tuples $(\sk, \Alg)$ where the output $\boldeta$ of $\Alg(\sk(x), \sk(y))$ is meant to estimate $d_X(x, y)$. The non-expansion guarantee remains the same---$\boldeta$ should not over-estimate $d_X(x,y)$ except with probability at most $\delta$. The bounded contraction guarantee is the one which differs; in worst-case sketching, the number $\boldeta$ should not fall below $d_X(x, y) / c$ with probability larger than $\delta$ for all $x, y \in X$ in a sketch with distortion $c$. The defining characteristic of Definition~\ref{def:avg-sketch} is the distribution $\mu$. Even though the algorithm ``knows'' $\mu$, it does not make any assumptions on it. In that regard, Theorem~\ref{thm:avg-sk-l-p} shows that, no matter what distribution, inputs being independently drawn are all that is needed. Note that, had $\mu$ been a distribution on pairs $(\bx,\by)$ over $X \times X$, then the definition would be equivalent to worst-case sketching via duality. For worst-case sketches, a theorem with space-distortion tradeoffs as in Theorem~\ref{thm:avg-sk-l-p} is known to be impossible, where $\Omega(d^{1-2/p} / c^2)$ bits are needed for a (worst-case) sketch of distortion $c$~\cite{BJKS04}.
}

\newcommand{\D}{\calD}

\ignore{ \newcommand{\M}{\calM}
\newcommand{\bsk}{\boldsymbol{\sk}}

\begin{definition}[Average-Case Embeddings]
    Given two metric spaces $(\calM,d_\M)$ and $(\N,d_\N)$, and an $n$-point dataset $P \subset \M$, we say a function $f: \M \to \N$ is an embedding from $\M$ into $\N$ with average distortion $\alpha$ with respect to dataset (discrete distribution) $P$ if
    \begin{itemize}
        \item For any $x,y \in \M$,
        \[d_\N(f(x),f(y))  \leq d_\M(x,y); \]
        \item It holds that
         \[  \Ex_{x,y\sim \D}[ d_\N(f(x),f(y))]\geq \frac{1}{\alpha} \cdot \Ex_{x,y\sim \D}[  d_\M(x,y)]\]
    \end{itemize}
\end{definition}

In many practical applications of metric embeddings, the quality of an embedding is measured in its average distortion. In theory, it is an alternative definition to describe the quality of an embedding or similarity between metric spaces. Several previous works studied average distortion embeddings \ldots

In the above context, (worst-case) distance estimation sketchings can be viewed as another alternative way of relaxation for worst-case embeddings, if one does not require the host space (the space of the sketches) to be a metric space and want small storage for each point. Oftentimes, the distance function for the sketches (or the decoder of the sketching) is more complicated. More formally, the basic setup asks us to design a (potentially randomized) sketching function $\bsk : X \to Y$ for a metric space $(X,d_X)$, so that for two points any $x,y \in X$, there exists a decoder that can estimate the distance $d_X(x,y)$ given only the sketches $\bsk(x), \bsk(y)$. Generally, $Y$ can be an arbitrary (potentially non-metric) domain, we want the size of storing $\bsk(x),\bsk(y) \in Y$ to be significantly smaller than that required of $x,y$. 

To study this problem, there is a natural communication model where Alice and Bob receive the points $x, y$ and send messages $m_A,m_B$ to a third party Charlie respectively; Charlie, acting as the decoder, outputs the estimator for $d_X(x,y)$ given $m_A,m_B$.
If an efficient protocol exists for the communication problem, then the function that generates the messages $m_A,m_B$ is exactly the sketching function as desired.
We focus on the problem where the metric space is $\mathbb{R}^d$ equipped with $\ell_p$ for some $p > 1$.

Unfortunately, \cite{lp-lower-bound} showed a lower bound on the message size for $\ell_p$ in the communication problem: For any $p>2$, to estimate $\|x-y\|_p$ within any constant factor requires $\Omega(d^{1-2/p})$-sized messages from Alice and Bob. Therefore, in order to get truly sublinear sketching for large $p$, we relax the problem by assuming that $x,y$ are not arbitrary, but rather independently \textit{drawn from some known distribution}.  We formally introduce the sketching problem studied in this paper below.

\paragraph{Average-Case $\ell_p$ Estimation Sketching Problem} Let $\D$ be a distribution supported on $\R^d$, which is fully known by Alice, Bob, and Charlie.
We also allow the parties access to public randomness.
Alice and Bob receive points $\bx,\by \sim D$, respectively, and send messages $m_A = \bsk(\bx,\D), m_B = \bsk(\by,\D)$ to Charlie, where $\bsk$ is a randomized function with implicit input random bits\footnote{Without loss of generality up to a factor 2 in communication complexity, we assume Alice and Bob use the same procedure $\bsk$ to generate their sketches.  If they used two separate procedures $\bsk_1,\bsk_2$, we could modify the protocol so they each run both $\bsk_1,\bsk_2$ and concatenate the outputs to produce the final message.}.
Charlie then must estimate $\|\bx-\by\|_p$ given $m_A,m_B$, and $\D$. When the distribution $D$ is clear given the context, we use $\bsk(.)$ without $D$ as an input.

We now introduce the definitions of average-case $\ell_p$ sketchings, which are closely related to average-case embeddings.
In our main definition, we require that the sketch is, in expectation, non-expanding for any pair of points, and that the expected sketch distance is at least a factor $1/\alpha$ of the true mean distance over $\D$.

\begin{definition} [Average-Case $\ell_p$ Estimation Sketching]
\label{def:acs-expectation}
    We say a randomized function $\bsk(s, \D)$ is an average-case $\ell_p$ estimation sketch with expected non-expansion for distribution $\D$ over $\mathbb{R}^d$ with distortion $\alpha\geq 1$ and error $\delta > 0$ if there exists a function $\Delta$ such that:
    \begin{enumerate}
        \item For any $x,y \in \mathbb{R}^d$, $\Ex_{\bsk}\left[ \Delta(\bsk(x),\bsk(y)) \right] \leq \|x-y\|_p $;
        \item $\mathbb{E}_{x,y\sim \D, \bsk}[\Delta(\bsk(x),\bsk(y))] \geq (1/\alpha)\mathbb{E}_{x,y\sim \D}\left[\|x-y\|_p\right]$.
    \end{enumerate}
\end{definition}

We are also able to obtain an \textit{instance-wise} probabilistic guarantee: when sampling $x,y\sim \D$, the sketched distance is at most a factor $\alpha$ smaller than the true distance with probability (over $\D$ and the sketch) at least $1-\delta$.

\begin{definition} [Distribution-Dependent $\ell_p$ Estimation Sketching]
\label{def:acs-probabilistic}
    We say a randomized function $\bsk(s,\D)$ is an average-case $\ell_p$ estimation sketch probabilistic non-expansion for distribution $\D$ over $\mathbb{R}^d$ with distortion $\alpha\geq 1$ and error $\delta > 0$ if there exists a function $\Delta$ such that:
    \begin{enumerate}
        \item For any $x,y \in \mathbb{R}^d$, $\pr_{\bsk}\left[ \Delta(\bsk(x),\bsk(y)) \leq \|x-y\|_p \right] \geq 1-\delta$;
        \item \(\displaystyle \pr_{x,y\sim \D, \bsk}\left[\Delta(\bsk(x),\bsk(y)) \geq \frac{1}{\alpha} \|x-y\|_p \right] \geq 1-\delta\).
    \end{enumerate}
\end{definition}

We obtain sketchings for both definitions by studying the decision version of average-case sketching, and then applying it repeatedly on candidate thresholds.

\ignore{paragraph{Average-Case $\ell_p$ Decision Sketching Problem} Let $\D$ be a distribution supported on $\R^d$ and $R > 0, \alpha > 1$ be given parameters.
$\D, R, \alpha$ are fully known by Alice, Bob, and Charlie.
We also allow the parties access to public randomness.
Alice and Bob receive points $\bx,\by \sim D$, respectively, and send messages $m_A = \bsk(\bx), m_B = \bsk(\by)$ to Charlie, where $\bsk$ is a randomized function with implicit input (public or private) random bits.
Charlie then must output $\close$ if $\|\bx-\by\|_p \leq R$ deterministically, and output $\far$  if $\|\bx-\by\|_p \geq \alpha \cdot R$ with certain probability over $\bx,\by \sim \D$ and $\bsk$, given $m_A,m_B$, and $\D$.
Charlie can output either $\close$ or $\far$ if $R < \|\bx-\by\|_p < \alpha \cdot R$.

We will mainly focus on the decision sketching problem with the following decision sketching definition.

\begin{definition} [Average-Case $\ell_p$ Decision Sketching]
\label{sec:acs-decision}
    We say a randomized function $\bsk(s,\D)$ is an average-case $\ell_p$ decision sketch for distribution $\D$ over $\mathbb{R}^d$, at threshold $R$, with distortion $\alpha\geq 1$ and errors $\delta_1,\delta_2 > 0$ if there exists a function $\Delta$ such that:
    \begin{enumerate}
        \item For any $x,y \in \mathbb{R}^d$, if $\|x-y\|_p \leq  R$, then it holds with probability $\delta_1$ that $\Delta(\bsk(x),\bsk(y)) = \close$;
        \item For any $\bx,\by \sim \D$, $\pr_{\bx,\by,\bsk}[\Delta(\bsk(\bx),\bsk(\by)) = \far \mid \|\bx-\by\|_p \geq \alpha \cdot R] \geq 1 - \delta_2$
    \end{enumerate}
\end{definition}

}

}

\section{Average-Distortion Sketching for \pdflp}\label{sec:ell-p}

\label{sec:ads-sketch}

In this section, we describe the average-distortion sketch for $([\Delta]^d, \ell_{p})$ and prove its correctness guarantees. Specifically, the proof of \cref{thm:ell-p-sketch} will follow from two steps. First, we show an average-distortion sketch for the ``single-scale'' decision version of the problem which succeeds with constant success probability.
Then, we show how the probability of incorrectly outputting $\far$ can be made arbitrarily small by ``boosting'' the error probability in \cref{subsec:proof-of-thm1}.\footnote{Note that, unlike worst-case sketches, it is not entirely clear how one boosts the success probability, since the far case is over a draw from a distribution $\mu$.}
Throughout the section, we think of preprocessing the distribution $\mu$ in the following manner: we let $z \in [\Delta]^d$ denote the vector which is the coordinate-wise median of the distribution $\mu$ ($z_i$ is the median of the marginal distribution of the $i$th coordinate), and we may consider the distribution which samples $\by \sim \mu$ and outputs $\by - z$.
This has the effect that the distribution is now supported on $\{-\Delta, \dots, \Delta\}^d$, and has coordinate-wise medians at the all-0's vector.

\begin{lemma}\label{lem:single-scale}
Let $\delta_1 = 1/32, \delta_2 = 1/8$.
For any $p \in [1, \infty)$, any distribution $\mu$ supported on $\{-\Delta, \dots, \Delta\}^d$ with median 0, and any $c \geq \minc$ (constant for $\delta_1,\delta_2$) and $r > 0$, there exists a distribution $\calD$ supported on tuples $(\sk, \Alg)$, where $\sk \colon [\Delta]^d \to \{0,1\}^s$ and $\Alg$ is an algorithm which outputs $\close$ or $\far$ which satisfies the following properties:
\begin{itemize}
\item \emph{\textbf{Non-Expansion}}: Any two points $x, y \in [\Delta]^d$ such that $\|x - y\|_{p} \leq r$ will have $\Alg(\sk(x), \sk(y)) = \close$ with probability at least $1 - \delta_1$ over the draw of the sketch.
\item \emph{\textbf{Bounded Contraction}}: We have that for any $x$ which satisfies $\|x\|_p \geq (c-1)r/2$,
\begin{displaymath}
	\Prx_{\substack{\by \sim \mu \\ (\sk, \Alg) \sim \calD}}\left[ \Alg(\sk(x),\sk(\by)) = \far \right] \geq \frac12 - \frac32 \cdot \delta_2
\end{displaymath}
\end{itemize}
The space complexity of the sketch is $2^{O(p/c)}$. Note that this complexity is independent of $d$ and $\Delta$ and \emph{only} a function of the ratio $p/c$.
\end{lemma}

We present the single-scale sketch, corresponding to Lemma~\ref{lem:single-scale} in Figure~\ref{fig:sketch}, and we note that it depends on the following setting of parameters. First, we let
\[ \sfD_1 = \ln(2/\delta_2)^{1/p} > 1 \qquad\text{and}\qquad \sfD_2 = 2 \cdot \sfD_1 > 2 \qquad\text{and}\qquad \sfD_3 = r / (\delta_1/6)^{1/p} > r, \]
which are fixed constants (for constants $\delta_1, \delta_2$). Then, we set the following parameters that are functions of $c$ and $p$:
\begin{itemize}
	\item \textbf{Number of Thresholds}: $\sfL = \floor*{c \cdot (\delta_1/6)^{1/p} / (4 \cdot \sfD_1) - 2} = \Theta(c)$. Since $c \geq \minc$, we have $\sfL \geq 2$.
	\item \textbf{Number of Useful Coordinates}: $\sfK = 2 \cdot \sfD_2^p / \delta_2 = 2^{\Theta(p)}$.%
	\item \textbf{Number of Stored Coordinates}: $\sfk = \ceil*{4 \cdot \sfK^{4/(\delta(\sfL-1))} + 2\ln(6 / \delta_1)} = 2^{\Theta(p/c)}$, where $\delta = \min\{\delta_1/6, \delta_2\}$. This parameter governs the space complexity of the sketch.
	\item \textbf{Hashing Universe Size}: $\sfU = 2 \cdot (6 + \sfk) / \delta_2$. %
\end{itemize}

We may assume $c$ is at most $O(p)$, as the average-distortion embedding already results in a constant-space sketch for approximation $O(p)$.

\begin{tcolorbox}[breakable, enhanced, colback=white]

\textbf{Single-Scale Sketch for $\ell_p$}. We will receive as input a vector $x \in [\Delta]^d$ to sketch with respect to the distribution $\mu$, as well as a desired distortion $c \geq \minc$ and scale $r > 0$, with parameters $\delta_1 = 1/32, \delta_2 = 1/8$.
We first describe the randomized map $\sk \colon [\Delta]^d \to \{0,1\}^s$ and then describe the algorithm $\Alg$. Let $z \in [\Delta]^{d}$ be the coordinate-wise median of the distribution $\mu$. 
By translating every vector by $-z$, we can assume the distribution $\mu$ is supported on $\{-\Delta, \dots ,\Delta\}^{d}$ such that the median of each coordinate is 0. We use public randomness of the draw $(\sk, \Alg) \sim \calD$ to sample:
\begin{itemize}
\item A random permutation $\bpi \colon [d] \to [d]$ which defines an order on the coordinates in $[d]$.
\item A random threshold $\bj \sim \{2, \dots, \sfL\}$.
\item Independent draws $\bu_1,\dots, \bu_d \sim \Exp(1)$. 
\item A random hash function $\bh_1 \colon [d] \to [\sfU]$ (used to store hashes of coordinates), and random hash function $\bh_2 \colon \Z \to [\sfU]$ (used to store hashes of rounded norms).
\end{itemize} 

\textbf{Sketching Map} $\sk$: Upon receiving the input $x \in \{ - \Delta, \dots, \Delta\}^d$:
\begin{enumerate}
\item \emph{Store} the hashes $\bh_2(\nu_x)$, $\bh_2(\nu_x+1)$, $\bh_2(\nu'_x)$, $\bh_2(\nu'_x+1)$ for $\nu_x = \ceil*{\|x\|_p / r}$ and $\nu'_x = \ceil*{ \|x\|_p / (\sfD_2 \cdot \sfD_3)}$. (If one is willing to use $O(\log d\Delta)$ bits, storing $\|x\|_p$ is sufficient for everything these hashes are used for in the output algorithm.) \label[skstep]{skstep:1}
\item For each $j \in \{0, \dots, \sfL \}$ and $\sigma \in \{-1,1\}$, define
\begin{displaymath}
	\tau(\nu, j) = \sfD_3 \cdot \ceil*{\frac{\nu}{\sfD_2 \cdot \sfD_3} + j} \qquad \text{and} \qquad G_{x,\bu}(t, \sigma) = \left\{i \in [d] :  \frac{x_i \cdot \sigma}{\bu_i^{1/p}} \geq t \right\}
\end{displaymath}
where $\tau(\nu, j)$ is the value of the $j$-th threshold, and $G_{x,\bu}(t, \sigma)$ is the set of coordinates which are at least the threshold value $t$.\label[skstep]{skstep:2}
 \item For each $\ell \in \{\bj-2, \bj-1, \bj\}$ and each $\sigma \in \{-1,1\}$, \emph{store} $\bh_1(i)$ for the first $\sfk$ coordinates $i \in G_{x,\bu}(\tau(\|x\|_p, \ell), \sigma)$ with respect to the permutation $\bpi$.
    Call this stored set of hashed indices $\bH_{x,\bu}(\tau(\|x\|_p, \ell),\sigma)$.
    \label[skstep]{skstep:3}
\item For each $\sigma \in \{-1,1\}$, \emph{store} $s_{x, \sigma} = 1$ if $|G_{x,\bu}(\tau(\|x\|_p,\bj-2),\sigma)| \leq (\sfk/4) \cdot |G_{x,\bu}(\tau(\|x\|_p,\bj),\sigma)|$ and $s_{x, \sigma} = 0$ otherwise.\label[skstep]{skstep:4}
\end{enumerate}

\textbf{Output Algorithm} $\Alg$: Upon receiving the input $\sk(x), \sk(y)$:
\begin{enumerate}

\item If $\{\bh_2(\nu_x), \bh_2(\nu_x + 1)\} \cap \{\bh_2(\nu_y), \bh_2(\nu_y + 1)\}$ is empty, output $\far$.
Otherwise, compute a value $\gamma$ in the following way.
If $\bh_2(\nu'_x) = \bh_2(\nu'_y)$ set $\gamma = 0$, if $\bh_2(\nu'_x+1) = \bh_2(\nu'_y)$ set $\gamma=1$, and if $\bh_2(\nu'_x) = \bh_2(\nu'_y+1)$ set $\gamma = -1$ (if none of these hold, output $\far$).
\label[step]{step:1}

\item By definition of $\gamma$ and $\tau$, we have $\tau(\|x\|_p, j-1) = \tau(\|y\|_p, j - 1 - \gamma)$ for any $j \in \{2, \dots, \sfL\}$, assuming there is no hash collision.
Now, if there exists a $\sigma \in \{-1,1\}$ such that $s_{x, \sigma} = 1$ and the first (with respect to $\bpi$) recorded value $\bh_1(i)$ from $\bH_{x,\bu}(\tau(\|x\|_p, \bj),\sigma)$ is not in $\bH_{y, \bu}(\tau(\|x\|_p, \bj-1), \sigma) = \bH_{y, \bu}(\tau(\|y\|_p, \bj-1-\gamma), \sigma)$ then output $\far$. \label[step]{step:2}

\item Otherwise, output $\close$. \label[step]{step:3} 

\end{enumerate}
\end{tcolorbox}
\begin{figure}[H]
    \vspace{-10pt}
    \caption{Single-Scale Sketch for $([\Delta]^d, \ell_{p})$} \label{fig:sketch}
\end{figure}

\begin{remark}
    An approximation of the coordinate-wise median suffices for our algorithm. 
    Specifically, for each coordinate $i\in [d]$, to produce the sketches one needs to know a value $m_i$ such that for $\bx\sim \mu$, $\pr[\bx_i \geq m_i] \geq q$ and $\pr[\bx_i \leq m_i] \geq q$, for $q=\Omega(1)$.
    With such a $q$, the only change to Lemma \ref{lem:single-scale} is that the probability in the bounded contraction case would then instead be $q-(3/2)\delta_2$.
    Finding all $m_i$ for $q$ close to $1/2$ (say $0.49$) can be done using only $O(\log d)$ samples from $\mu$.
\end{remark}

\begin{lemma}\label{lem:close}
Suppose $x, y \in \{ - \Delta, \dots, \Delta\}^d$ with $\|x-y\|_p \leq r$. Then, the probability over $(\sk, \Alg) \sim \calD$, that $\Alg(\sk(x), \sk(y))$ outputs $\far$ is at most $\delta_1$.
\end{lemma}

Before proving \cref{lem:close}, we first prove the following lemma.
\begin{lemma}\label{lem:bad-coord}
    Consider fixed $x, y \in [\Delta]^d$ such that $\|x - y\|_p \leq r$.
    Then the probability that there exists a coordinate $i \in [d]$ such that $|x_i - y_i| \geq \sfD_3 \cdot \bu_i^{1/p}$ is at most $\delta_1/6$.
\end{lemma}
\begin{proof}
    Using the definition of $\sfD_3$, and the fact that $\bu_i \sim \Exp(1)$, we have
    \begin{align*}
        \Prx_{\bu_1,\dots, \bu_d \sim \Exp(1)}\left[ \exists i \in [d] : \frac{|x_i - y_i|}{\bu_i^{1/p}} \geq \frac{r}{(\delta_1/6)^{1/p}} \right] &= 1 - \prod_{i=1}^d \Prx_{\bu_i \sim \Exp(1)}\left[ \frac{|x_i - y_i|}{\bu_i^{1/p}} \leq \frac{r}{(\delta_1/6)^{1/p}} \right] \\
                &= 1 - \exp\left( - \frac{\|x-y\|_p^p}{r^p} \cdot \frac{\delta_1}{6} \right) \leq \frac{\delta_1}{6} \cdot \left( \frac{\|x-y\|_p}{r}\right)^p \leq \delta_1/6.
    \end{align*}
\end{proof}

\begin{proof}[Proof of \cref{lem:close}]
There are several ways that $\Alg(\sk(x), \sk(y))$ (hereafter referred to as ``the algorithm'') may output $\far$ when $\|x - y\|_p \leq r$. 
\begin{enumerate}
    \item The first comes from \cref{step:1} of the output algorithm $\Alg$: when the hashes $\{ \bh_2(\nu_x), \bh_2(\nu_x+1)\}$ and $\{ \bh_2(\nu_y), \bh_2(\nu_y+1)\}$ do not intersect. If $\|x-y\|_p \leq r$, then $|\nu_x - \nu_y| \leq 1$, and therefore the sets always intersect and the algorithm never outputs $\far$ in this step.
    Similarly, if the hashes $\{ \bh_2(\nu_x'), \bh_2(\nu_x'+1)\}$ and $\{ \bh_2(\nu_y'), \bh_2(\nu_y'+1)\}$ do not intersect, then $|\nu_x'-\nu_y'| > 1$; this is impossible when $\|x-y\|_p \leq r$ as $\sfD_2\cdot \sfD_3 > r$.
    \item 
    Consider $j$, a fixed setting of the random variable $\bj \sim \{2, \dots, \sfL\}$, and consider the case where $s_{x, \sigma} = 1$ for some $\sigma \in \{-1,1\}$, and define sets $\bX_{-2} = G_{x, \bu}(\tau(\|x\|_p, j-2), \sigma)$ and $\bX_0 = G_{x, \bu}(\tau(\|x\|_p, j), \sigma)$ and $\bY_{-1} = G_{y, \bu}(\tau(\|x\|_p, j-1), \sigma)$.
    So, by definition of $s_{x,\sigma}$, we have the following inequality:
    \begin{align}
        |\bX_{-2}| \leq (\sfk/4) \cdot |\bX_0| \enspace. \label{eq:size-constraint}
    \end{align}
    There are two further subcases:
    \begin{enumerate}
        \item $|\bY_{-1}| > (\sfk / 4) \cdot |\bX_0|$.  \label[case]{case:1}
        \item $|\bY_{-1}| \leq (\sfk / 4) \cdot |\bX_0|$. In this case, can only output $\far$ when $\|x - y\|_p \leq r$ if:
        \begin{enumerate}
            \item the first coordinate $i_1 \in \bX_0$ with respect to $\bpi$ is not in $\bY_{-1}$; or \label[case]{case:2}
            \item the first coordinate $i_1 \in \bX_0$ with respect to $\bpi$ is in $\bY_{-1}$, but there are $\sfk$ other coordinates $i_2 \in \bY_{-1}$ such that $\bpi(i_2) < \bpi(i_1)$ (and thus $\sk(y)$ does not include a hash of $i_1$). \label[case]{case:3}
        \end{enumerate}
    \end{enumerate}
\end{enumerate}

We consider the probability of \cref{case:1,case:2,case:3} one by one.
In all of these cases, the algorithm can output $\far$ only if there exists $\sigma \in \{-1, 1\}$ such that $s_{x, \sigma} = 1$ (\cref{eq:size-constraint} is satisfied).

Firstly, consider \cref{case:1}, where $|\bY_{-1}| > (\sfk/4) \cdot |\bX_0|$.
Since \cref{eq:size-constraint} is satisfied, it must further be the case that $|\bY_{-1}| > |\bX_{-2}|$.
Therefore, there exists a coordinate $i \in [d]$ such that $y_i \sigma / \bu_i^{1/p} \geq \tau(\|x\|_p, j-1)$ and $x_i \sigma / \bu_i^{1/p} < \tau(\|x\|_p, j-2)$.
Rearranging and using the definition of $\tau$, we see that there exists an $i \in [d]$ such that $|x_i - y_i| > \sfD_3 \cdot \bu_i^{1/p}$.
However, by \cref{lem:bad-coord}, the probability that this occurs is at most $\delta_1/6$.
Union bounding over $\sigma \in \{-1,1\}$, the probability of this case occurring is at most $\delta_1/3$.

\cref{case:2} is similar, where $|\bY_{-1}| \leq |\bX_0|$ is satisfied, but the first coordinate $i \in \bX_0$ with respect to $\bpi$ does not lie in $\bY_{-1}$.
In order for this to happen, \smash{$x_i \sigma / \bu_i^{1/p} \geq \tau(\|x\|_p, \bj)$} and \smash{$y_i \sigma / \bu_i^{1/p} < \tau(\|x\|_p, \bj-1)$}.
However, by \cref{lem:bad-coord}, the probability that this occurs is at most $\delta_1/6$.
Union bounding over $\sigma \in \{-1,1\}$, the probability of this case occurring is at most $\delta_1/3$.

Finally, for \cref{case:3}, consider any fixed setting $u$ and $j$ of random variables $\bu$ and $\bj$ (and thus also fixing sets $X_{-2}, Y_{-1}, X_0$).
This case occurs whenever there exist at least $\sfk$ coordinates $j \in Y_{-1}$ such that $\bpi(j) < \bpi(\bi)$, where $\bi$ is the first coordinate of $X_0$ with respect to $\bpi$, even though the sets satisfy (\ref{eq:size-constraint}). In order to upper-bound this probability, consider the following mechanism for sampling the permutation $\bpi$:
\begin{enumerate}
    \item First, sample a permutation $\bpi_1 \colon X_0 \to [|X_0|]$ that defines an order of the elements lying only in $X_0$.
    \item Second, sample a map $\bpi_2 \colon Y_{-1} \setminus X_0 \to \{0, \dots, |X_0|\}$, which defines the relative position of each element of $Y_{-1} \setminus X_0$ with respect to $X_0$. That is, any element $z \in Y_{-1} \setminus X_0$ will lie between the $\bpi_2(z)$-th element of $X_0$ and the $(\bpi_2(z)+1)$-th element of $X_0$.
    \item Finally, we draw permutations for each $\bpi_2^{-1}(\ell)$ for all $\ell \in \{0,\dots, |X_0|\}$, and order the remaining elements in $[d] \setminus (X_0 \cup Y_{-1})$.
\end{enumerate}
The above procedure generates a uniformly random permutation $\bpi$, and the random variable $\bi \in X_0$ is fixed after sampling $\bpi_1$ in the first step. Then, in order for this third case to occur, we must have $|\bpi^{-1}_2(0)| \geq \sfk$; however, $|\bpi_2^{-1}(0)|$ is a sum of $|Y_{-1} \setminus X_0|$ independent Bernoulli random variables, each of which occurs with probability $1/(|X_0| + 1)$. Thus, $|\bpi_2^{-1}(0)|$ has expectation at most $|Y_{-1}| / |X_0 | \leq \sfk / 4$, and by Bernstein's inequality, the probability that $|\bpi^{-1}_2(0)|$ is at least $\sfk$ is at most $e^{-\sfk/2}$.
The probability of the last case occurring is at most $\delta_1/3$ as long as $\sfk \geq 2\ln(6 / \delta_1)$, so we can union bound over $\sigma \in \{-1,1\}$.

Therefore, the total probability of any of the three cases occurring is at most $\delta_1$.
\end{proof}

It remains to lower bound the probability that the protocol outputs $\far$ when the input $x$ satisfies $\|x\|_p \geq (c-1)r/2$ and $\by \sim \mu$ is drawn from the distribution.
We first show the following lemma, which will allow us to fix the randomness of $\bu$.
In particular, for any vector $x \in \{-\Delta, \dotsc, \Delta\}^d$, with high probability over the choice of the embedding vector $\bu$, there is at least one coordinate $i$ with mapped value $\bx_i' = x_i/\bu_i^{1/p}$ such that $|\bx_i'|\geq \|x\|_p/\sfD_1$, but at most $\sfK = 2^{O(p)}$ coordinates $i$ where $|\bx_i'|\geq \|x\|_p/\sfD_2$.

\begin{lemma}\label{lem:good-event}
For any vector $x \in \{-\Delta, \dots, \Delta\}^d$, 
\begin{align*}
&\Prx_{\bu_1, \dots, \bu_d \sim \Exp(1)}\left[ 1 \leq \left| \bigcup_{\sigma \in \{-1,1\}} G_{x,\bu}\left( \frac{\|x\|_p}{\sfD_1}, \sigma\right) \right| \leq \left| \bigcup_{\sigma \in \{-1,1\}} G_{x,\bu}\left(\frac{\|x\|_p}{\sfD_2}, \sigma\right) \right| \leq \sfK  \right] \\
&\qquad\qquad\qquad \geq 
 1 - \exp\left( - \sfD_1^p \right) - \frac{\sfD_2^p}{\sfK} \geq 1 - \delta_2.
\end{align*}
\end{lemma}

\begin{proof}
For simplicity in the notation, let $\bx' \in \R^d$ denote the vector given by $\bx_i' = x_i / \bu_i^{1/p}$ for all $i \in [d]$. Then, we will lower bound the above probability by (i) upper bounding the probability over $\bu$ that $\max_{i \in [d]} |\bx_i'|$ is less than $\|x\|_p / \sfD_1$, and then (ii) upper bounding the probability over $\bu$ that the number of coordinates $i \in [d]$ where $|\bx_i'|$ is at least $\|x\|_p / \sfD_2$ is larger than $\sfK$.
The total probability of failure is the sum of these two probabilities by the union bound, which then gives us the desired bound.

For (i), we have
\begin{align*}
\Prx_{\bu}\left[ \max_{i \in [d]} |\bx_i'| < \frac{\|x\|_p}{\sfD_1} \right] = \Prx_{\bu}\left[\forall i \in [d]: \bu_i > \frac{|x_i|^p}{(\|x\|_p / \sfD_1)^{p}}\right] = \prod_{i\in [d]} \Prx_{\bu_i}\left[\bu_i > \frac{|x_i|^p}{(\|x\|_p / \sfD_1)^{p}}\right] = \exp\left( - \sfD_1^p\right).
\end{align*}
To upper bound (ii), we first compute the expected number of coordinates with $|\bx_i'| \geq \|x\|_p / \sfD_2$ and apply Markov's inequality.
By linearity of expectation, the expected number of such coordinates $i$ is given by
\begin{align*}
\sum_{i=1}^d \Prx_{\bu_i} \left[\frac{|x_i|}{\bu_i^{1/p}} \geq \frac{ \|x\|_p }{\sfD_2} \right] = \sum_{i=1}^d \Prx_{\bu_i}\left[ \bu_i \leq \frac{\sfD_2^p |x_i|^p}{\|x\|_p^p}\right] \leq \sum_{i=1}^d \left(1 - \exp\left(-\frac{\sfD_2^p |x_i|^p}{\|x\|_p^p} \right) \right) \leq \sfD_2^p,
\end{align*}
where the first inequality holds by the cumulative probability distribution of an exponential random variable, and the second inequality follows from $1 - \exp(-\alpha) \leq \alpha$.
By Markov's inequality, the probability that more than $\sfK$ coordinates $i \in [d]$ satisfy $|\bx_i'| \geq \|x\|_p / \sfD_2$ is at most $\sfD_2^p/\sfK$ concluding the upper bound proof of (ii).
\end{proof}

From the above lemma, we define the following indicator variables\footnote{Throughout this section, we will define many such events $\textcolor{\eventlinkcolor}{\calE_i}$.  For the reader's convenience, each of these events are clickable links back to their definition.} which will help us lower bound the probability that the sketch outputs $\far$.
These events depend only on the randomness over $\bu_1,\dots, \bu_d \sim \Exp(1)$, which will allow our analysis to then consider any fixed setting of $x$ and $u$ satisfying these events:
\begin{enumerate}[label={\normalfont \textbullet}]
    \item We let $\Elink{1}(x)$ denote the event, which only depends on $x$, that $\|x\|_p \geq (c-1)r/2$.\label{event:1}
\item Fix any $x$ where $\Elink{1}(x)$ holds. We let $\Elink{2}(x, u)$ denote the event, which depends only on $x$ and $u$ that the event of Lemma~\ref{lem:good-event} holds.
    \label{event:2}
\end{enumerate}
\begin{remark}[Fixing value of $\sigma$]
\label{rem:fix-sigma}
    For a fixed $x$ and $u$ where $\Elink{1}(x)$ and $\Elink{2}(x,u)$ hold, let $\sigma \in \{-1,1\}$ denote the sign so that $\max_{i \in [d]} \sigma \cdot x_i' = \max_{i \in [d]}|x_i'|$. Note that such a setting of $\sigma$ implies that the set $G_{x,u}(t, \sigma)$ is non-empty for all $t \in [0,\tau(\|x\|_p, \sfL)]$, since $\tau(\|x\|_p, \sfL) \leq \|x\|_p / \sfD_1$:
\begin{displaymath}
    \tau(\|x\|_p, \sfL) \leq \frac{\|x\|_p}{\sfD_2} + (\sfL + 1) \frac{r}{(\delta_1/6)^{1/p}} \leq \frac{\|x\|_p}{2\sfD_1} + \paren*{\frac{c \cdot (\delta_1/6)^{1/p}}{4\sfD_1} - 1} \frac{r}{(\delta_1/6)^{1/p}} \leq \frac{\|x\|_p}{2\sfD_1} + \frac{(c-1)r}{4\sfD_1} \leq \frac{\|x\|_p}{\sfD_1}.
\end{displaymath}
\end{remark}

Using the definition of $\Elink{1}(x)$ and $\Elink{2}(x, u)$, we state the following main technical lemma, and show how to invoke it to prove \cref{lem:single-scale}.

\begin{lemma}\label{lem:prob-close-small}
Consider any fixed setting of $x \in \{-\Delta, \dots, \Delta\}^d$, $u \in \R_{\geq 0}^d$ and $\sigma \in \{-1,1\}$ where $\Elink{1}(x)$ and $\Elink{2}(x, u)$ hold. Then,
\begin{align*}
\Prx_{\substack{\by \sim \mu \\ \bpi, \bh_1, \bh_2}}\left[ \Alg(\sk(x), \sk(\by)) = \close \right] \leq \frac12 + \delta_2.
\end{align*}

\end{lemma}

\begin{proof}[Proof of Lemma~\ref{lem:single-scale} assuming Lemma~\ref{lem:prob-close-small}] 
Lemma~\ref{lem:close} implies that any two vectors $x, y$ within distance at most $r$ will output $\close$ with probability at least $1-\delta_1$, giving us the non-expansion guarantee of the sketch. It suffices to prove the bounded contraction. Consider any fixed setting of $x$ where $\Elink{1}(x)$ occurs.  Then,
\begin{align*}
& \Prx_{\substack{\by \sim \mu \\ (\sk, \Alg) \sim \calD}}\left[ \Alg(\sk(x),\sk(\by)) = \far \right] \\
&\qquad \geq \Prx_{\substack{\by \sim \mu \\ \bu, \bpi, \bh_1, \bh_2}}\left[ \Alg(\sk(x),\sk(\by)) = \far
\mid \Elink{2}(x, \bu)\right] \nonumber \times \Prx_{\bu }\left[ \Elink{2}(x, \bu) \right] \nonumber \\
&\qquad \geq \Prx_{\substack{\by \sim \mu \\ \bu, \bpi, \bh_1, \bh_2}}\left[ \Alg(\sk(x),\sk(\by)) = \far 
\mid \Elink{2}(x, \bu)\right] \nonumber \times (1 - \delta_2) \enspace, \label{eq:todo}
\end{align*}
where the second inequality follows from Lemma~\ref{lem:good-event}. Finally, Lemma~\ref{lem:prob-close-small}, which upper bounds the probability that $\Alg(\sk(x), \sk(\by))$ outputs $\close$ for any $u$ where $\Elink{1}(x)$ and $\Elink{2}(x,u)$ hold, may equivalently be used to lower bound the probability the sketch output $\far$ conditioned on $\Elink{2}(x,\bu)$ for fixed $x$ with $\Elink{1}(x)$.
Putting the two inequalities together,
\begin{align*}
\Prx_{\substack{\by \sim \mu \\ (\sk, \Alg) \sim \D}}\left[ \Alg(\sk(x),\sk(\by)) = \far \right] \geq \paren*{\frac12 - \delta_2}\paren*{1-\delta_2} \geq \frac12 - \frac32\delta_2.
\end{align*}
For the space complexity, note that we are storing:
\begin{itemize}
    \item The hash value of $\bh_2(\nu_x), \bh_2(\nu_x + 1), \bh_2(\nu'_x), \bh_2(\nu'_x + 1)$, using $O(\log \sfU)$ bits.
    \item For each $\ell \in \{\bj - 2, \bj - 1, \bj\}$ and $\sigma \in \{-1, 1\}$, $\sfk$ many hashes $\bh_1(i)$.
\end{itemize}
Thus, the total space complexity is
\[ O(\log \sfU) + O\left( \sfk \cdot \log \sfU \right) = 2^{O(p/c)}. \qedhere\]
\end{proof}

\subsection{Proof of Lemma~\ref{lem:prob-close-small}}
\label{sec:main-lemma-avg}

The proof will proceed by first describing a sequence of events, and claiming that, for any deterministic choice of $y$ and $\pi$, the fact that $\Alg(\sk(x),\sk(y))$ outputs $\close$ (using the permutation $\pi$ and $u$) implies that one of the events must have held. Then, we individually upper bound the probability that each of the events holds, which will give the desired bound.

\begin{definition}
Consider any fixed setting of $x, y \in \{-\Delta, \dots, \Delta\}^d$, and any fixed setting of $u$, $\pi$, $h_1, h_2$ and $\sigma$. Suppose that $\Elink{1}(x)$ and $\Elink{2}(x,u)$ both hold, and let $\nu = \|x\|_p$. We consider any $j \in \{2, \dots, \sfL\}$, and define the variable $i_{\mathsf{xmin}} \in [d]$ as the index given by the first element of $G_{x,u}(\tau(\nu,j), \sigma)$ with respect to the permutation $\pi$.
(Notice that the variable $i_{\mathsf{xmin}}$ is well-defined since $G_{x,u}(\tau(\nu,j), \sigma)$ is non-empty, and is a function of $x, j, \sigma$ and $\pi$, but importantly, not $y$.)

Now, we define events which will help us understand the reasons that $\Alg(\sk(x), \sk(y))$ outputs $\close$. 
Importantly, the following events do not depend on $\pi$:
\begin{enumerate}[label={\normalfont \textbullet}]
\item We let $\Elink{3}(x, y)$ denote the event that $\left| \|x\|_p - \|y\|_p \right| \leq 2r$.\label{event:3}
\item We let $\Elink{4}(x, y, u, j, \sigma)$ denote the event that $\Elink{3}(x, y)$ holds and that $\nu'_x + \gamma = \nu'_y$ (recall that $\nu'_x = \ceil*{\|x\|_p / (\sfD_2 \cdot \sfD_3)}$, and $\gamma$ is defined in \cref{step:1} of the output algorithm $\Alg$).
Additionally, $s_{x, \sigma} = 1$ and therefore \cref{eq:size-constraint} holds. \label{event:4}
\end{enumerate}
We now define the events which do depend on $\pi$:
\begin{enumerate}[label={\normalfont \textbullet}]
\item We let $\Elink{5}(x, y, u,j,\sigma, \pi)$ denote the event that $\sign(x_{i_{\mathsf{xmin}}}) \neq \sign(y_{i_{\mathsf{xmin}}})$\footnote{We use the standard convention for the $\sign$ function: $\sign(x) = 1$ if $x > 0$, $\sign(x) = -1$ if $x < 0$ and $\sign(x) = 0$ if $x = 0$.}, which implies $i_{\mathsf{xmin}} \notin G_{y, u}(t, \sigma)$ for any $t > 0$ (recall that $i_{\mathsf{xmin}}$ is a function of $x,u,j,\pi$ and $\sigma$, but does not depend on $y$). \label{event:5}
\item We let $\Elink{6}(x,y,u,j,\sigma, \pi, h_1)$ denote the event that $\Elink{5}(x,y,u,j,\sigma,\pi)$ holds, and in addition, each of the first $\sfk$ coordinates $t \in G_{y,u}(\tau(\nu, j-1), \sigma)$ (which do not include $i_{\mathsf{xmin}}$) satisfy $h_1(i_{\mathsf{xmin}}) \neq h_1(t)$. \label{event:6}

\end{enumerate}
\end{definition}

\begin{figure}[ht]
\centering
\quad\quad
    \begin{tikzpicture}
        \begin{scope}[every node/.style={circle,line width=1pt,draw}, minimum size=0.85cm, scale=0.95, inner sep=0pt]
            \node (a) at (0,0) {$\Elink{3}$};
            \node (b) at (-1.25,-2) {(i)};
            \node (c) at (2,0) {$\Elink{4}$};
            \node (d) at (0.75,-2) {(ii)};
            \node (e) at (4,0) {$\Elink{5}$};
            \node (f) at (2.75,-2) {(iii)};
            \node (g) at (6,0) {$\Elink{6}$};
            \node (h) at (4.75,-2) {(iv)};
            \node (i) at (7.25, -2) {\footnotesize{$\far$}};
        \end{scope}

        \begin{scope}[style={line width=0.9pt}]
            \draw[->] (a) -- (b) node[midway, above left=-2pt] {F};
            \draw[->] (a) -- (c) node[midway, above] {T};
            \draw[->] (c) -- (d) node[midway, above left=-2pt] {F};
            \draw[->] (c) -- (e) node[midway, above] {T};
            \draw[->] (e) -- (f) node[midway, above left=-2pt] {F};
            \draw[->] (e) -- (g) node[midway, above] {T};
            \draw[->] (g) -- (h) node[midway, above left=-2pt] {F};
            \draw[->] (g) -- (i) node[midway, above right=-2pt] {T};
        \end{scope}
    \end{tikzpicture}
    \caption{Accompanying diagram for the proof of Lemma~\ref{lem:bad-events}. The events are labeled as nodes, leading to, either the cases considered in Lemma~\ref{lem:bad-events}, or two nodes labeled $\far$ where the sketch will output $\far$. Edges are labelled ``T'' or ``F'', corresponding to whether the events in nodes hold (in the case ``T'') or do not hold (in the case ``F'').} \label{fig:diagram}
\end{figure}

\begin{lemma}\label{lem:bad-events}
Consider a fixed setting of $x, y, u, \pi,\sigma, h_1, h_2$, as well as $j \in \{2, \dots, \sfL\}$, and suppose $\Elink{1}(x)$ and $\Elink{2}(x,u)$ hold. If $\Alg(\sk(x),\sk(y))$ outputs $\close$ using $u$ and $\pi$, then at least one of the following must be true:
\begin{itemize}
\item[(i)] The event $\Elink{3}(x,y)$ does not hold, but the hashes $\{ h_2(\nu_x), h_2(\nu_x+1)\}$ and $\{ h_2(\nu_y), h_2(\nu_y+ 1)\}$ intersect.
\item[(ii)] The event $\Elink{3}(x,y)$ holds, but $\Elink{4}(x, y, u, j, \sigma)$ does not hold. 
\item[(iii)] The event $\Elink{5}(x, y, u, j, \sigma, \pi)$ does not hold.
\item[(iv)] The event $\Elink{5}(x, y, u, j, \sigma, \pi)$ holds, but $\Elink{6}(x, y, u, j, \sigma, \pi, h_1)$ does not hold.
\end{itemize}
\end{lemma}

\begin{proof}
We prove the above lemma by showing the contrapositive, which comes down to the following cases:
\begin{enumerate}
\item Suppose that $\Elink{3}(x,y)$ does not hold and the hashes do not intersect, then $\Alg$ outputs $\far$ directly by \cref{step:1} of the output algorithm $\Alg$.
\item Suppose that $\Elink{3}(x,y), \Elink{4}(x, y, u, j, \sigma)$, $\Elink{5}(x,y,u,j,\sigma,\pi)$, and $\Elink{6}(x,y,u,j,\sigma,\pi,h_1)$ hold. Then, $\Alg$ outputs $\far$.
We claim $\Alg$ can determine that $h_1(i_{\mathsf{xmin}})$, which is encoded in $\sk(x)$, is not among the first $\sfk$ values of $h_1(j)$ for $j \in G_{y,u}(\tau(\nu,j-1), \sigma)$.
This occurs for the following reason.
By definition of $i_{\mathsf{xmin}}$, it is the first element of $G_{x, u}(\tau(\nu,j), \sigma)$ and thus $h_1(i_{\mathsf{xmin}})$ is stored in $\sk(x)$.
Since $\Elink{4}(x, y, u, j, \sigma)$ occurs, there is some $j' \in \{j-2, j-1, j\}$ where $\tau(\|y\|_p, j') = \tau(\nu, j-1)$, so $\sk(y)$ stores the hashes of the first $\sfk$ elements of $G_{y,u}(\tau(\nu, j-1), \sigma)$, but $i_{\mathsf{xmin}} \notin G_{y,u}(\tau(\nu,j-1), \sigma)$ because $\Elink{5}(x, y, u,j,\sigma,\pi)$ implies $\sign(x_{i_{\mathsf{xmin}}}) \neq \sign(y_{i_{\mathsf{xmin}}})$; finally, by $\Elink{6}(x,y,u,j,\sigma,\pi)$, the first $k$ hashes of $G_{y,u}(\tau(v, j-1), \sigma)$ do not equal $h_1(i_{\mathsf{xmin}})$, so the algorithm outputs $\far$ by \cref{step:2} of the output algorithm $\Alg$.\qedhere
\end{enumerate}
\end{proof}

From \cref{lem:bad-events}, we consider fixed $x$ and $u$ where $\Elink{1}(x)$ and $\Elink{2}(x,u)$ hold, and fixed $\sigma$ according to \cref{rem:fix-sigma}. 
We sample $\by \sim \mu$, $\bpi$ as a random permutation of $[d]$, the hash functions $\bh_1$ and $\bh_2$, and a random threshold $\bj \sim \{2, \dots, \sfL\}$.
We then upper bound the probability that $\Alg(\sk(x), \sk(\by))$ outputs $\close$ by upper-bounding all of the probabilistic events in Lemma~\ref{lem:bad-events} (and applying a union bound).
\begin{align}
& \Prx_{\substack{\by \sim \mu \\ \bpi, \bh_1, \bh_2}}\left[ \Alg(\sk(x), \sk(\by)) = \close\right] \notag \\
    & \qquad \leq \Prx_{\by, \bh_2}\left[\begin{array}{c} \neg \Elink{3}(x, \by) \wedge \\ \{ \bh_2(\nu_x), \bh_2(\nu_x+1) \} \cap \{ \bh_2(\nu_y), \bh_2(\nu_y+1)\} \neq \emptyset \end{array} \right] + \Prx_{\by, \bj}\left[  \begin{array}{c} \Elink{3}(x,\by) \wedge \\ \neg \Elink{4}(x, \by, u, \bj, \sigma) \end{array} \right] \label{eq:hah1}\\
    & \qquad \qquad + \Prx_{\substack{\by, \bj \\ \bpi}}\left[ \neg \Elink{5}(x,\by, u, \bj,\sigma, \bpi)\right] + \Prx_{\substack{\by,\bj,\\\bpi,\bh_1}}\left[ \begin{array}{c} \Elink{5}(x,\by,u,\bj,\sigma,\bpi) \wedge \\\neg \Elink{6}(x,\by,u,\bj,\sigma, \bpi, \bh_1)\end{array}\right] \label{eq:hah2}
\end{align}
In the above inequality, we can refer to the terms and their corresponding nodes of Figure~\ref{fig:diagram}.
The first and second terms in (\ref{eq:hah1}) correspond to (i) and (ii) respectively.
The first and second terms in (\ref{eq:hah2}) correspond to (iii) and (iv) respectively. 
In the subsequent lemmas, we upper bounds the values of these probabilities, and substituting these lemmas will upper bound the above probability by
\begin{align*}
    \frac{4}{\sfU} + \paren*{\frac{2}{\sfU} + \frac{\delta_2}{2}} + \frac12 + \frac{\sfk}{\sfU} \leq \frac12 + \delta_2, \nonumber
\end{align*}
by the setting of $\sfk$ and $\sfU$.

\begin{lemma}
    Fix any $x, y \in \{-\Delta, \dots, \Delta\}^d$ where $\Elink{3}(x, y)$ does not occur, i.e., $\|x\|_p$ and $\|y\|_p$ differ by at least $2r$. Then, the probability that there is a non-empty intersection among $\bh_2(\nu_x), \bh_2(\nu_x+1)$ and $\bh_2(\nu_y)$ and $\bh_2(\nu_y+1)$ is at most $4 / \sfU$, and therefore the first term of (\ref{eq:hah1}) is at most $4/\sfU$.
\end{lemma}

\begin{proof}
    Recall that $\nu_x = \ceil*{\|x\|_p / r}$ and $\nu_y = \ceil*{\|y\|_p / r}$. 
    The lemma follows from a straight-forward union bound.
    Since $\Elink{3}(x, y)$ does not occur, $\|x\|_p$ and $\|y\|_p$ differ by at least $2r$.
    Therefore, $|\nu_x - \nu_y| \geq 2$.
    Thus, $\nu_x, \nu_{x}+1$, $\nu_y$ and $\nu_{y} + 1$ are all distinct.
    Since there are $4$ pairs of possible collisions, and each one occurs with probability at most $1/\sfU$, the probability of any collision is at most $4/\sfU$.
\end{proof}

\begin{lemma}
\label{lem:calE4}
Consider any fixed $x,y \in \{-\Delta, \dots, \Delta\}^d$ and $u$ such that $\Elink{1}(x), \Elink{2}(x,u)$ and $\Elink{3}(x,y)$ occur. 
If $\sigma$ is fixed according to \cref{rem:fix-sigma}, then $\Prx_{\bj}\left[\neg \Elink{4}(x, y, u, \bj, \sigma)\right] = 2/\sfU + \delta_2/2$ and therefore the second term of (\ref{eq:hah2}) is $2/\sfU + \delta_2/2$.
\end{lemma}
\begin{proof}
Recall that we defined $\nu'_x = \ceil*{\|x\|/(\sfD_2 \cdot \sfD_3)}$ and $\nu'_y = \ceil*{\|y\|/(\sfD_2 \cdot \sfD_3)}$.
Since $\Elink{3}(x, y)$ occurs, $\sfD_2 > 2$, and $\sfD_3 > r$, we have 
\begin{align*}
   \left|\|x\|_p - \|y\|_p \right| \leq 2r \implies \left| \ceil*{\|x\|_p / (\sfD_2 \cdot \sfD_3)} - \ceil*{\|y\|_p / (\sfD_2 \cdot \sfD_3)} \right| \leq 1 \implies \left| \nu'_x - \nu'_y \right| \leq 1 \enspace.
\end{align*}
Since $\nu'_x$ and $\nu'_y$ differ by at most $1$, we have $\gamma \in \{-1, 0, 1\}$, where $\gamma$ is defined in \cref{step:1} of the output algorithm $\Alg$, where it was set such that $\bh_2(\nu'_x + \gamma) = \bh_2(\nu'_y)$.
As long as $\bh_2(\nu'_x) \neq \bh_2(\nu'_x + 1)$ and $\bh_2(\nu'_y) \neq \bh_2(\nu'_y + 1)$, this implies that $\nu'_x + \gamma = \nu'_y$.
The probability of either of these collisions occurring is at most $2/\sfU$.

Now, $s_{x, \sigma} = 1$ exactly when
\begin{displaymath}
    |G_{x,u}(\tau(\|x\|_p,\bj-2),\sigma)| \leq (\sfk/4) \cdot |G_{x,u}(\tau(\|x\|_p,\bj),\sigma)| \enspace.
\end{displaymath}

Since we fixed $x$ and $u$ such that $\Elink{2}(x, u)$ occurs, we have $|G_{x,u}(\tau(\nu, 0), \sigma)| \leq \sfK$.
Moreover, we fixed $\sigma$ such that $|G_{x,u}(\tau(\nu, \sfL), \sigma)| \geq 1$.
We also have that the number of $j \in \{2, \dots, \sfL\}$ such that $|G_{x,u}(\tau(\nu, j-2), \sigma)| > (\sfk/4) \cdot |G_{x,u}(\tau(\nu, j), \sigma)|$ is at most $\delta_2 \cdot (\sfL-1) / 2$.
The reason for this as follows.
If the number of such $j$ were more than $\delta_2 \cdot (\sfL-1) / 2$, one of the two must occur:
\begin{itemize}
    \item Where $E$ is the set of even $j \in \{2, \dots, \sfL\}$, there are at least $\delta_2 \cdot (\sfL-1)/4$ indices $j \in E$ satisfying $|G_{x,u}(\tau(\nu, j-2), \sigma)| > \sfK^{4/(\delta_2(\sfL-1))} \cdot |G_{x,u}(\tau(\nu,j), \sigma)|$ (recall that $\sfk \geq 4 \cdot \sfK^{4/(\delta_2(\sfL-1))}$).
    \item Or, where $O$ is the set of odd $j \in \{2, \dots, \sfL \}$, there are at least $\delta_2 \cdot (\sfL-1)/4$ indices  $j \in O$ satisfying $|G_{x,u}(\tau(\nu, j-2), \sigma)| > \sfK^{4/(\delta_2(\sfL-1))} \cdot |G_{x,u}(\tau(\nu,j), \sigma)|$. 
\end{itemize}
Assuming $\sfL$ is even, we can rewrite $|G_{x,u}(\tau(\nu, 0), \sigma)|$ as
\begin{align*}
|G_{x,u}(\tau(\nu, 0), \sigma)| &= \prod_{j\in E} \dfrac{|G_{x,u}(\tau(\nu, j-2), \sigma)|}{|G_{x,u}(\tau(\nu, j), \sigma)|} \cdot |G_{x,u}(\tau(\nu, \sfL), \sigma)| \\
& = \left(\dfrac{|G_{x, u}(\tau(\nu, 0), \sigma)|}{|G_{x,u}(\tau(\nu,1), \sigma)|} \right) \prod_{j \in O} \left(\dfrac{|G_{x,u}(\tau(\nu, j-2), \sigma)|}{|G_{x,u}(\tau(\nu, j), \sigma)|} \right) \cdot |G_{x,u}(\tau(\nu, \sfL-1), \sigma)| \enspace,
\end{align*}
Since $\Elink{2}(x, u)$ occurs, $|G_{x,u}(\tau(\nu, \sfL-1), \sigma)| \geq 1$ and $|G_{x,u}(\tau(\nu, \sfL), \sigma)| \geq 1$.
Additionally, all terms in the product are at least one, since sets are nested and always contain at least one element.
Thus, in either of the two cases, we obtain
\[ |G_{x,u}(\tau(\nu, 0), \sigma)| > \left( \sfK^{4/(\delta_2(\sfL-1))} \right)^{\delta_2(\sfL-1)/4} = \sfK, \]
contradicting $\Elink{2}(x,u)$.
Thus, a draw of $\bj \sim \{2, \dots, \sfL\}$ fails to satisfy \cref{eq:size-constraint} when it falls within a set of size $\delta_2 (\sfL - 1) / 2$ out of $\sfL - 1$ choices, in which case $s_{x, \sigma} = 0$ and $\Elink{4}(x, y, u, \bj, \sigma)$ does not hold.
\end{proof}

\begin{lemma}
\label{lem:calE7}
Consider any fixed setting of $x, u$ such that $\Elink{1}(x)$ and $\Elink{2}(x,u)$ occur, and any $j, \sigma$ and $\pi$. Then, we have 
\begin{align*}
\Prx_{\by}\left[ \neg \Elink{5}(x, \by, u, j, \sigma, \pi) \right] \leq \frac12
\end{align*}
and therefore, the first term of (\ref{eq:hah2}) is at most $1/2$.
\end{lemma}
\begin{proof}
$\Elink{5}(x, \by, u, j, \sigma, \pi)$ does not occur if $\sign(x_{i_{\mathsf{xmin}}}) = \sign(\by_{i_{\mathsf{xmin}}})$.
Note that $i_{\mathsf{xmin}}$ is a function of $x, j, \sigma$ and $\pi$, but not $\by$: for fixed values of $x, j, \sigma, \pi$, the value of $\sign(x_{i_{\mathsf{xmin}}})$ is fixed. Firstly, by definition of $i_{\mathsf{xmin}}$, $x_{i_{\mathsf{xmin}}} \neq 0$: this is because $i_{\mathsf{xmin}}$ is defined for $j \geq 1$, and $\tau(\nu, j) > 0$ if $j \geq 1$.
Now, if $\sign(x_{i_{\mathsf{xmin}}}) = 1$, then $\Prx_{\by}\brac*{\neg \Elink{5}(x, \by, u, j, \sigma, \pi)} = \Prx_{\by}\brac*{\sign(\by_{i_{\mathsf{xmin}}}) = 1} \leq \frac12$, since the median of the marginal distribution under $\mu$ of each coordinate is 0.
Similarly, if $\sign(x_{i_{\mathsf{xmin}}}) = -1$, $\Prx_{\by}\brac*{\neg \Elink{5}(x, \by, u, j, \sigma, \pi)} = \Prx_{\by}\brac*{\sign(\by_{i_{\mathsf{xmin}}}) = -1} \leq \frac12$ as desired.
\end{proof}

\begin{lemma}
    Consider any fixed setting of $x, y, u, j, \sigma, \pi$ such that $\Elink{5}(x, y,u,j,\sigma,\pi)$ occurs. 
    Then the probability over $\bh_1$ that $\Elink{6}(x,y,u,j,\sigma,\pi,\bh_1)$ does not occur is at most $\sfk / \sfU$.
    Therefore, the second term of (\ref{eq:hah2}) is at most $\sfk/\sfU$.
\end{lemma}

\begin{proof}
    This follows from the fact that we are upper bounding the probability that there is a hash collision from a hash function which maps to a universe of size $\sfU$ among $\sfk$ possible pairs. 
\end{proof}

\subsection{Proof of Theorem~\ref{thm:ell-p-sketch}}\label{subsec:proof-of-thm1}

We begin by first describing a lemma which uses multiple instances of the sketch of \cref{lem:single-scale}, and is useful in the proof of \cref{thm:ell-p-sketch}.
This lemma is different from \cref{lem:single-scale} in just one way: the probability of non-expansion is not treated as a constant, and can be parameterized by $\delta_0$ which is counted in the space complexity of the sketch.

\begin{lemma}\label{lem:boost-prob-of-non-expansion}
Consider any $\delta_0 \in (0, 1)$, $r > 0$, $c$ at least some sufficiently large fixed constant, $p \geq 1$, and a distribution $\mu$ over $\{-\Delta, \dots, \Delta\}^d$ such that the median of each marginal is 0. Then, there exists a sketching algorithm $(\sk, \Alg) \sim \calD^{\circ}$ satisfying the following properties:
\begin{itemize}
	\item \emph{\textbf{Non-Expansion}}: Given any two points $x, y \in \{-\Delta, \dots, \Delta\}^d$ such that $\|x - y\|_{p} \leq r$, $\Alg(\sk(x), \sk(y)) = \close$ with probability at least $1 - \delta_0$ over the draw of the sketch.
	\item \emph{\textbf{Bounded Contraction}}: For a fixed $x \in \{-\Delta, \dots, \Delta\}^d$ such that $\|x\|_p \geq (c-1)r/2$, we have
\begin{displaymath}
	\Prx_{\substack{\by \sim \mu \\ (\sk, \Alg) \sim \calD^{\circ}}}\left[ \Alg(\sk(x),\sk(\by)) = \far\right] \geq \frac14 
\end{displaymath}
\end{itemize}
The space complexity of the sketch is $2^{O(p/c)}\log(1/\delta_0)$.
\end{lemma}
\begin{proof}
We first describe the sketch and algorithm.
Let $(\sk_1, \Alg_1), \dots, (\sk_T, \Alg_T) \sim \calD$ be $T=\lceil 512\ln(1/\delta_0) \rceil$ independent samples of the sketch from \cref{lem:single-scale}, each instantiated with constants $\delta_1 = 1/32, \delta_2 = 1/8$.
Define the following distribution $\calD^{\circ}$ over $(\sk, \Alg)$, where $\sk(x) = (\sk_1(x), \dots, \sk_T(x))$ is the concatenation of the $T$ individual sketches, and $\Alg(\sk(x), \sk(y))$ outputs $\far$ if at least $T/16$ of $\Alg_i(\sk_i(x), \sk_i(y))$ output $\far$, and otherwise outputs $\close$.

We now prove the non-expansion bound, so consider two arbitrary fixed points $x, y \in \{-\Delta, \dots, \Delta\}^d$ such that $\|x - y\|_{p} \leq r$.
We define $T$ indicator random variables: for each $i \in [T]$, $\bX_i$ is 0 if $\Alg_i(\sk_i(x), \sk_i(y))$ returns $\close$ and 1 otherwise.
By \cref{lem:single-scale},
\begin{displaymath}
	\Prx_{(\sk_i, \Alg_i) \sim \calD}[\bX_i = 1] \leq \delta_1
\end{displaymath}
for all $i \in [T]$.
Let $\bX = \sum_{i=1}^T \bX_i$. We have $\Ex_{(\sk, \Alg) \sim \calD^{\circ}}[\bX] \leq \delta_1 \cdot T = T/32$.
Using the Hoeffding bound, we get 
\begin{displaymath}
    \Prx_{(\sk, \Alg) \sim \calD^{\circ}}\left[ \Alg'(\sk'(x), \sk'(y)) = \far \right] = \Prx[\bX \geq T/16] \leq \Prx[\bX - \Ex[\bX] \geq T/32] \leq \exp(-T/512) \leq \delta_0.
\end{displaymath}
We are able to apply the Hoeffding bound because the $\bX_i$ are independent, since the sketches are drawn independently from $\calD$.

Now, we prove the bounded contraction bound.
Consider an arbitrary fixed point $x \in \{-\Delta, \dots, \Delta\}^d$ such that $\|x\|_p \geq (c-1)r/2$.
We define indicator random variables $\bY_i$ for each $i \in [d]$ such that $\bY_i=1$ if $\Alg_i(\sk_i(x), \sk_i(\by))=\close$ and 0 otherwise.
Note that $\bY_i$ depends on the both the randomness of the sketch $(\sk_i, \Alg_i)\sim \calD$ and $\by \sim \mu$.
Let $\bY = \sum_{i=1}^T \bY_i$. Using the bounded-contraction bound of each instance of $(\sk_i, \Alg_i)$, we get 
\begin{displaymath}
	\Ex_{\substack{\by \sim \mu \\ (\sk, \Alg) \sim \calD^{\circ}}}[\bY] \leq T \cdot \Prx_{\substack{\by \sim \mu \\ (\sk_1, \Alg_1) \sim \calD}}\left[ \Alg(\sk_1(x),\sk_1(\by)) = \close \right] \leq T \cdot \paren*{\frac12 + \frac32\delta_2} = T \cdot \frac{11}{16}
\end{displaymath}
using $\delta_2 = 1/8$.
The variables $\bY_i$ are not necessarily independent of each other, as they depend on the randomness of $\by \sim \mu$; nonetheless, by Markov's inequality,
\begin{align*}
    \Prx_{\substack{\by \sim \mu \\ (\sk, \Alg) \sim \calD^{\circ}}}\brac*{\Alg(\sk(x), \sk(\by)) = \close} & = \Prx_{\substack{\by \sim \mu \\ (\sk, \Alg) \sim \calD^{\circ}}}\brac*{\bY \geq T \cdot \frac{15}{16}}\\
	& \leq \Prx_{\substack{\by \sim \mu \\ (\sk, \Alg) \sim \calD^{\circ}}}\brac*{\bY \geq \Ex[\bY] \cdot \frac{16}{11} \cdot \frac{15}{16}} \leq \frac{11}{15} \leq \frac34
\end{align*}
as desired.
\end{proof}
\begin{proof}[Proof of \cref{thm:ell-p-sketch}]
We use the bounded contraction bound from \cref{lem:boost-prob-of-non-expansion} to lower bound the probability of the algorithm outputting $\far$ for a pair of $\bx, \by \sim \mu$:

\begin{align*}
	& \Prx_{\substack{\bx, \by \sim \mu \\ (\sk, \Alg) \sim \calD^{\circ}}}\left[ \Alg(\sk(\bx), \sk(\by)) = \far \right] \\
	& \qquad = \Prx_{\substack{\bx, \by \sim \mu \\ (\sk, \Alg) \sim \calD^{\circ}}}\left[ \Alg(\sk(\bx), \sk(\by)) = \far \mid \|\bx\|_p \geq (c-1)r/2 \right] \times \Prx_{\bx \sim \mu}\left[ \|\bx\|_p \geq (c-1)r/2 \right] \\
    & \qquad \qquad + \Prx_{\substack{\bx, \by \sim \mu \\ (\sk, \Alg) \sim \calD^{\circ}}}\left[ \Alg(\sk(\bx), \sk(\by)) = \far, \|\bx\|_p < (c-1)r/2 \right] \\
    & \qquad \geq \frac14 \cdot \Prx_{\bx \sim \mu}\left[ \|\bx\|_p \geq (c-1)r/2 \right] + \Prx_{\substack{\bx, \by \sim \mu \\ (\sk, \Alg) \sim \calD^{\circ}}}\left[ \Alg(\sk(\bx), \sk(\by)) = \far, \|\bx\|_p < (c-1)r/2\right] \\
    & \qquad \geq \frac14 \cdot \Prx_{\bx \sim \mu}\left[ \|\bx\|_p \geq (c-1)r/2 \right] + \Prx_{\substack{\bx, \by \sim \mu \\ (\sk, \Alg) \sim \calD^{\circ}}}\left[ \Alg(\sk(\bx), \sk(\by)) = \far, \|\by\|_p \geq (c+3)r/2, \|\bx\|_p < (c-1)r/2\right] \\
	& \qquad \geq \frac14 \cdot \Prx_{\bx \sim \mu}\left[ \|\bx\|_p \geq (c-1)r/2 \right] + \lp 1 - \frac{4}{\sfU} \rp \cdot \Prx_{\bx,\by \sim \mu}\left[ \|\by\|_p \geq (c+3)r/2, \|\bx\|_p < (c-1)r/2\right] \\
    & \qquad \geq \frac14 \cdot \Prx_{\bx \sim \mu}\left[ \|\bx\|_p \geq (c-1)r/2 \right] + \lp 1 - \frac{4}{\sfU} \rp\cdot \Prx_{\bx,\by \sim \mu}\left[ \|\bx - \by\|_p \geq (c+1)r, \|\bx\|_p < (c-1)r/2 \right] \\
    & \qquad \geq \frac14 \cdot \Prx_{\bx,\by \sim \mu}\left[ \|\bx - \by\|_p \geq (c+1)r, \|\bx\|_p \geq (c-1)r/2 \right] + \frac14 \cdot \Prx_{\bx,\by \sim \mu}\left[ \|\bx - \by\|_p \geq (c+1)r, \|\bx\|_p < (c-1)r/2 \right] \\
	& \qquad = \frac14 \cdot \Prx_{\bx,\by \sim \mu}\left[ \|\bx-\by\|_p \geq (c+1)r \right]
\end{align*}
where
\begin{enumerate}
    \item the first inequality follows from \cref{lem:boost-prob-of-non-expansion}, since $\Prx\left[ \Alg(\sk(\bx), \sk(\by)) = \far \mid \|\bx\|_p \geq (c-1)r/2 \right] \geq 1/4$,
    \item the second inequality follows from adding an additional event in the joint probability,
    \item the third inequality follows from the fact that 
    \begin{displaymath}
        \Prx_{\bx,\by \sim \mu}\left[ \Alg(\sk(\bx), \sk(\by)) = \far \mid \|\by\|_p \geq (c+3)r/2, \|\bx\|_p < (c-1)r/2\right] \geq 1 - \frac{4}{\sfU}
    \end{displaymath}
    because there are at least 2 integers that lie between $\nu_x$ and $\nu_y$ if $\|\by\|_p \geq (c+3)r/2$ and $\|\bx\|_p < (c-1)r/2$, and therefore \cref{step:1} of the output algorithm in \cref{fig:sketch} fails to return $\far$ only if there is a hash collision, which occurs with probability at most $4/\sfU$,
    \item the fourth inequality follows from the triangle inequality: if both $\|\bx - \by\|_p \geq (c+1)r$ and $\|\bx\|_p < (c-1)r/2$ occur, then $\|\by\|_p \geq (c+3)r/2$ occurs,
    \item the fifth inequality follows from adding an additional event in the joint probability, and because $1 - 4/\sfU \geq 1/4$.
\end{enumerate}
So, if we instead sample $(\sk,\Alg)$ from the distribution of Lemma \ref{lem:boost-prob-of-non-expansion} with distorition parameter $c-1$, we obtain
\begin{equation}
\label{eq:far-vs-cr}
    \Prx \left[ \Alg(\sk(\bx), \sk(\by)) = \far \right] \geq \frac14 \cdot \Prx \left[ \|\bx-\by\|_p \geq cr \right].
\end{equation}

	The sketch for \cref{thm:ell-p-sketch} proceeds by concatenating $O(\log(d\Delta))$ many sketches from \cref{lem:boost-prob-of-non-expansion} at possible values of $r = 2^w$.
    Specifically, for any $w$, let $\calD_w$ be the distribution over sketches from Lemma \ref{lem:boost-prob-of-non-expansion} with distortion parameter $c-1$, $r=2^w$, and $\delta_0 = 1/d\Delta$;
    further define $W = \{\floor*{-\log_2(c)}, \dots, \ceil*{\log_2(d\Delta/c)}\}$.
	The distribution $\calD$ over $(\sk, \Alg)$ is then given by sampling $(\sk_w, \Alg_w) \sim \calD_w$ for all $w \in W$:
	\begin{itemize}
	\item The map $\sk \colon X \to \{0,1\}^s$ concatenates the maps $\sk_w$ for each $w \in W$, which is what increases the space complexity by a multiplicative $O(\log(d\Delta))$ factor.
    \item The algorithm $\Alg$, upon receiving $\sk(x)$ and $\sk(y)$, executes $\Alg_w(\sk_w(x), \sk_w(y))$ for every $w$ and outputs $2^{\bw^{*}-2}$, where $\bw^*$ is the largest $w\in W$ with $\Alg_{w}(\sk_{w}(x), \sk_{w}(y))=\far$.
    If all algorithms output $\close$, output $0$.
	\end{itemize}

    For non-expansion, fix $x,y \in [\Delta]^{d}$ with $x\neq y$, and let $v=\floor*{\log_2(\|x-y\|_p)}$ be the largest $w$ such that $2^{w} \leq \|x-y\|_p$.
    So, by construction, for all $w \geq v$, we have $\Alg_w(\sk_w(x), \sk_w(y)) = \close$ with probability at least $1-\delta_0$ by the non-expansion condition of Lemma \ref{lem:boost-prob-of-non-expansion}.

    Then, by definition of expectation and the distribution $\calD$, we have
    \begin{align*}
        \Ex_{(\sk,\Alg)\sim \calD}\left[ \Alg(\sk(x),\sk(y)) \right] & \leq \sum_{w = \floor*{-\log_2(c)}}^{\ceil*{\log_2(d\Delta/c)}} 2^{w-2} \cdot \Prx_{(\sk_w,\Alg_w)\sim \calD_w}\left[ \Alg_w(\sk_w(x),\sk_w(y)) = \far \right] \\
        &\leq \sum_{w = \floor*{-\log_2(c)}}^{v-1} 2^{w-2}  + \sum_{w = v}^{\ceil*{\log_2(d\Delta/c)}} 2^{w-2} \cdot \delta_0 \\
        &\leq \left(2^{v-2} + \frac{1}{2}\right) + \frac{1}{2}\left(\delta_0 \cdot \frac{d\Delta}{2}\right)\\
        &= 2^{v-2} + \frac{3}{4}
    \end{align*}
    as $\delta_0 = 1/d\Delta$ and $c\geq 2$.
    When $x\neq y$, $v\geq 0$ as $\|x-y\|_p \geq 1$, and so $2^{v-2} + 3/4 \leq 2^{v} \leq \|x-y\|_p$, as desired.
    When $x=y$, we output $\close$ at every scale since $\Alg_w(\sk_w(x), \sk_w(x))=\close$ always\footnote{This follows immediatly from the definition of the decoding algorithm $\Alg_w$: whenever $\sk_w(x)=\sk_w(y)$, the algorithm will output $\close$. 
    }, and so $\Alg(\sk(x),\sk(y)) = 0 = \|x-y\|_p$.
	
	For proving bounded average contraction, the geometrically increasing weights $2^w$ allow us to lower bound the expected value outputted by the sketch, and apply \cref{lem:boost-prob-of-non-expansion}.
    Since the potential scales $2^w$ are geometrically increasing, for any subset $S \subseteq W$, $\max_{w\in S}2^w \geq (1/2) \sum_{w\in S} 2^w$.
    So,
	\begin{align*}
	\Ex_{\substack{\bx, \by \sim \mu \\ (\sk, \Alg) \sim \calD}}\left[ \Alg(\sk(\bx), \sk(\by)) \right] & \geq \frac{1}{2}\sum_{w = \floor*{-\log_2(c)}}^{\ceil*{\log_2(d\Delta/c)}} 2^{w-2} \Prx_{\substack{\bx,\by \sim \mu \\ (\sk_w, \Alg_w) \sim \calD_w}}\left[ \Alg_w(\sk_w(\bx), \sk_w(\by)) = \far\right] \\
	& \geq \frac{1}{2}\sum_{w = \floor*{-\log_2(c)}}^{\ceil*{\log_2(d\Delta/c)}} 2^{w-2} \cdot \frac14 \cdot \Prx_{\bx, \by \sim \mu} \left[ \| \bx-\by\|_p \geq c \cdot 2^w \right] & [\cref{eq:far-vs-cr}]\\
	& \geq \frac{1}{64c} \cdot \sum_{w = \floor*{-\log_2(c)}}^{\ceil*{\log_2(d\Delta/c)}} c \cdot 2^{w+1} \cdot \Prx_{\bx, \by \sim \mu} \left[ \| \bx-\by\|_p \in [c \cdot 2^w, c \cdot 2^{w+1}) \right] \\
	& \geq \frac{1}{64c} \cdot \Ex_{\bx, \by \sim \mu}\left[ \|\bx-\by\|_p \cdot \ind\{ \|\bx-\by\|_p \in [1, 2d\Delta] \} \right] \\
	& = \frac{1}{64c} \cdot \Ex_{\bx,\by\sim \mu}\left[ \|\bx-\by\|_p \right]
	\end{align*}
    where the last equality follows from the fact that all non-zero $\ell_p$ distances in $\{-\Delta, \dots, \Delta\}^d$ at least 1.

    So, when invoking Lemma \ref{lem:boost-prob-of-non-expansion} with distortion parameter $c'=c/64$, we obtain the desired bounded contraction bound from Theorem \ref{thm:ell-p-sketch}.
    So, replacing distortion parameter $c$ with $c'=c/64$ in the above sketch and analysis, we obtain the desired bounded contraction bound from Theorem \ref{thm:ell-p-sketch}.

	The space complexity of the sketch is
	\begin{displaymath}
		  2^{O(p/c)}\log\left(\frac{1}{\delta_0}\right) \cdot O(\log(d\Delta)) = 2^{O(p/c)} \cdot \log^2(d\Delta) \qedhere
	\end{displaymath}
	\end{proof}

\subsection{Asymmetric Sketch}
\label{sec:asymmetric-sketch}
The single-scale sketch described in the previous section can be made ``asymmetric'' with almost no changes.
Let $\mu$ be a distribution on $\{-\Delta, \dots, \Delta\}^d$, and let $x,y \sim \mu$ be two independent samples from $\mu$.
In this setting, we allow the sketches for $x$ and $y$ to be computed by different algorithms and be of different sizes.
Specifically, the sketch size of $y$ remains the same (which is $2^{O(p/c)}$), but the sketch size of $x$ is reduced to only $O(p/c)$.
The output algorithm of \cref{fig:sketch} uses minimal information from $\sk(x)$ to determine whether $x$ and $y$ are $\close$ or $\far$.
This information only requires $O(p/c)$ bits to store, and therefore we can modify the sketch of $x$ to only store this information, thereby achieving our desired space complexity.
This asymmetric sketch with different sizes for the two inputs is critical in our construction of the approximate nearest-neighbor data structure (see \cref{sec:applications}).

We formalize this notion of an asymmetric sketch in the following corollary of Lemma \ref{lem:boost-prob-of-non-expansion}.

\begin{corollary}[Lemma \ref{lem:boost-prob-of-non-expansion} with asymmetric sketch sizes]
    \label{corr:asymmetric-sketch}
    Consider any $\delta_0 \in (0, 1)$, $r > 0$, $c$ at least some sufficiently large fixed constant, $p \geq 1$, and a distribution $\mu$ over $\{-\Delta, \dots, \Delta\}^d$ such that the median of each marginal is 0.
    Then, there exists a distribution $\calD$ supported on tuples $(\sk_A, \sk_B, \Alg)$ where $\sk_A: [\Delta]^{d}\to \{0,1\}^{s_A}$, $\sk_B: [\Delta]^{d}\to \{0,1\}^{s_B}$ and $\Alg$ is an algorithm which outputs $\close$ or $\far$.
    Moreover, $\Alg$ satisfies the following properties:
    \begin{itemize}
        \item \emph{\textbf{Non-Expansion}}: Given any two points $x, y \in \{-\Delta, \dots, \Delta\}^d$ such that $\|x - y\|_{p} \leq r$, $\Alg(\sk_A(x), \sk_B(y)) = \close$ with probability at least $1 - \delta_0$ over the draw of the sketches.
        \item \emph{\textbf{Bounded Contraction}}: For a fixed $x \in \{-\Delta, \dots, \Delta\}^d$ such that $\|x\|_p \geq (c-1)r/2$, we have
    \begin{displaymath}
        \Prx_{\substack{\by \sim \mu \\ (\sk_A, \sk_B, \Alg) \sim \calD^{\circ}}}\left[ \Alg(\sk_A(x),\sk_B(\by)) = \far\right] \geq \frac14.
    \end{displaymath}
    \end{itemize}
    The space complexity of $\sk_A$ is $\displaystyle O\left(\frac{p}{c} \cdot \log(1/\delta_0)\right)$ and the space complexity of $\sk_B$ is $2^{O(p/c)}\log(1/\delta_0)$.
\end{corollary}

\begin{proof}
    We first show the case with $\delta_0 = 1/32$, i.e.~the analogous version of Lemma \ref{lem:single-scale}; the result then follows by the same argument used in the proof of Lemma \ref{lem:boost-prob-of-non-expansion}.

    The sketch is nearly identical to Figure \ref{fig:sketch}, with the only change being that $\sk_A$ stores only the first coordinate in Step \ref{skstep:3}, rather than the first $\sfk = 2^{\Theta(p/c)}$.
    Specifically:
    \begin{itemize}
        \item $\sk_A(x)$ is constructed in the same way as $\sk(x)$ from Figure \ref{fig:sketch}, except for Step \ref{skstep:3}.
            Instead of Step \ref{skstep:3}, for each $\sigma \in \{-1,1\}$ store $\bh_1(i)$ for the first coordinate $i\in G_{x,\bu}(\tau(\|x\|_p,\bj),\sigma)$ with respect to the permutation $\bpi$ (rather than storing the hashes of the first $\sfk$ coordinates as in Figure \ref{fig:sketch}).
        \item $\sk_B(y)$ is constructed identically to $\sk(y)$ from Figure \ref{fig:sketch}.
        \item Given sketches $\sk_A(x),\sk_B(y)$, the output algorithm $\Alg$ is the same as in Figure \ref{fig:sketch}.
    \end{itemize}
    The space complexity of $\sk_B$ is immediate from Lemma \ref{lem:single-scale}.
    $\sk_A(x)$ stores $O(1)$ hashes, each of which are integers in $[\sfU]$.
    Since $\sfU = 2^{O(p/c)}$, it follows that the size of $\sk_A(x)$ is $O(\log \sfU) = O(p/c)$, as desired.

    Notice that the output algorithm of Figure \ref{fig:sketch} only uses the first element of $\bH_{x,\bu}(\tau(\|x\|_p,\bj),\sigma)$ for each $\sigma\in \{-1,1\}$, even if more are stored, and does not use inspect any elements potentially stored in $\bH_{x,\bu}(\tau(\|x\|_p,\ell),\sigma)$ for any $\ell\neq\bj$.
    So, removing these from the stored sketch (which are the modifications made to obtain $\sk_A(x)$) does not affect the output algorithm.
    Thus for any $x,y\in \{-\Delta,\ldots,\Delta\}$, since $\sk_B(y)=\sk(y)$ we have that $\Alg(\sk_A(x),\sk_B(y)) = \Alg(\sk(A),\sk(y))$ and the result follows from Lemma \ref{lem:single-scale}.\qedhere

\end{proof}

Corollary \ref{corr:asymmetric-sketch} then immediately implies the following weaker sketching scheme where the output algorithm has complete access to one point.
We use this corollary in our nearest-neighbor data structure in Section \ref{sec:applications}.

\begin{corollary}[One-sided Sketch]
    \label{corr:one-way}
    Consider any $\delta_0 \in (0, 1)$, $r > 0$, $c$ at least some sufficiently large fixed constant, $p \geq 1$, and a distribution $\mu$ over $\{-\Delta, \dots, \Delta\}^d$ such that the median of each marginal is 0.
    Then, there exists a distribution $\calD$ supported on tuples $(\sk, \Alg)$ where $\sk: [\Delta]^{d}\to \{0,1\}^{s}$ and $\Alg$ is an algorithm which outputs $\close$ or $\far$ and satisfies the following properties:
    \begin{itemize}
        \item \emph{\textbf{Non-Expansion}}: Given any two points $x, y \in \{-\Delta, \dots, \Delta\}^d$ such that $\|x - y\|_{p} \leq r$, $\Alg(\sk(x), y) = \close$ with probability at least $1 - \delta_0$ over the draw of the sketch $\sk$.
        \item \emph{\textbf{Bounded Contraction}}: For a fixed $x \in \{-\Delta, \dots, \Delta\}^d$ such that $\|x\|_p \geq (c-1)r/2$, we have
    \begin{displaymath}
        \Prx_{\substack{\by \sim \mu \\ (\sk, \Alg) \sim \calD}}\left[ \Alg(\sk(x),\by) = \far\right] \geq \frac14.
    \end{displaymath}
    \end{itemize}
    The space complexity of the sketch is $\displaystyle O\left(\frac{p}{c}\cdot \log(1/\delta_0)\right)$.
\end{corollary}

\section{Application to Nearest Neighbor Search in \pdflp}
\label{sec:applications}

We now prove Theorem~\ref{thm:nns-p}, which applies
Corollary~\ref{corr:one-way}
in designing algorithms for approximate nearest neighbor search under the $\ell_p$ norm. In particular, it suffices to solve the $(c,r)$-approximate near neighbor problem to solve the $c$-approximate nearest neighbor problem (up to additional polylogarithmic factors in space and query time)~\cite{HIM12}. 
\begin{definition}[$(c,r)$-Approximate Near Neighbor]
    For $r > 0$ and $c \geq 1$, the $(c,r)$-approximate near neighbor problem for $(\R^d, \ell_p)$ is the following data structure problem:
    \begin{itemize}
        \item \emph{\textbf{Preprocessing}}: A dataset $X$ of $n$ vectors in $\R^d$ is preprocessed into a data structure.
        \item \emph{\textbf{Query}}: A query is specified by a point $q \in \R^d$ such that $\exists x \in X$ with $\|x- q\|_p \leq r$, and the data structure should output a point $x' \in X$ with $\|x'- q\| \leq cr$.
    \end{itemize}
    For randomized data structure, any fixed dataset $X$ and query $q$ should produce a correct answer on query $q$ with high probability over internal randomness of the data structure.
\end{definition}

\ignore{We prove the following theorem, which improves on the approximation from $O(p)$, to $c_0 (c+1)$ for any $c \geq 1$, and $c_0$ being a fixed universal constant. 
\begin{theorem}
    For any $p \in [1, \infty)$ and $c \geq 1$ and any $r, \eps > 0$, there is a data structure for $(c,r)$-approximate near neighbor over $\ell_p$ with the following guarantees:
    \begin{itemize}
        \item \emph{\textbf{Query Time}}: The time to execute a single query is $O(n^{\eps} \log n) \cdot \left(d + \poly(cp \cdot 2^{p/c}) \right)$.
        \item \emph{\textbf{Space and Preprocessing Time}}: The data structure preprocesses a dataset in $d\cdot n^{\poly(cp \cdot 2^{p/c}) \cdot \log(1/\eps)}$ time and build a data structure of that much space.
    \end{itemize}
\end{theorem}}
The data structure we construct will be a direct (recursive) application of the single-scale average-distortion sketch of Corollary \ref{corr:one-way} as a type of locality-sensitive hash function (our reduction will very much follow~\cite{JWZ24}, for data-dependent locality-sensitive hashing).
In particular, we present a ``core'' data structure which will recursively apply the sketch $O(\log n)$ times, and generate a rooted tree of depth $O(\log n)$ and arity $2^{s}$, where $s$ is the sketch size.
The query then follows a single root-to-leaf path, and succeeds at finding an approximate near neighbor with probability at least $n^{-\eps}$. The ``core'' data structure is then repeated $O(n^{\eps})$ times to boost the success probability. 
Since the preprocessing step has access to the entire dataset, we are able to use the (exponentially smaller) ``one-way'' sketches of Corollary \ref{corr:one-way} rather than the full single-scale sketch of Lemma \ref{lem:boost-prob-of-non-expansion}.

\begin{remark}[Discretizing Data]
    In nearest neighbor search, one generally considers vectors in $\R^d$ without necessarily specifying the bit-complexity. Since we use sketches for $(\{-\Delta,\dots, \Delta\}^d, \ell_p)$, we discretize all dataset vectors and queries to be in $\{-\Delta, \dots, \Delta\}^d$. We emphasize that this is without loss of generality because our single-scale average-distortion sketches (Lemma~\ref{lem:boost-prob-of-non-expansion} and Corollary \ref{corr:one-way}) have no quantitative dependence on $\Delta$. For $(c, r)$-approximate near neighbor, discretizing by rounding coordinates to the nearest integer multiple of $\eps r / d$ incurs at most an additive $\eps r$ factor to distances, and then up to translation and re-scaling, one can work with a discrete metric $[\Delta]^d$ for a large enough $\Delta$.
\end{remark}

\begin{figure}[H]
\begin{framed}

\textsc{Core-Preprocess}$(X, k)$. \\

\textbf{Input}: A dataset $X \subset [\Delta]^d$ and an integer $k$. We assume that the scale $r > 0$, approximation $c > 1$, and parameter $\eps > 0$ is fixed. \\
\textbf{Output}: The pointer to a data structure node $v$.

\begin{itemize}
    \item Initialize a data structure node $v$.
    \item If $k = 0$, store $X$ in $v.\textit{data}$ and return $v$.
    \item If $k > 0$, perform the following:
    \begin{enumerate}
        \item Let $\mu$ be the uniform distribution over $X$, $m \in [\Delta]^d$ denote the coordinate-wise median, and $x \in X$ be any point within $(c+1)r/2$ of $m$ if one exists. Store $m$ in $v.\textit{med}$ and $x$ in $v.\textit{close}$, and let $\tilde{\mu}$ be the centered distribution $\bz - m$ for $\bz \sim \mu$ (whose median is the all-$0$'s vector).
        \item Initialize a sketch $(\sk, \Alg)$ for $\tilde{\mu}$ from Corollary \ref{corr:one-way} with scale $r$, approximation $c$, and failure probability $\delta_0 = \eps$, which sketches $\{-\Delta, \dots, \Delta\}^d$ to $\{0,1\}^s$. Store the randomness of $\sk$ in $v.\textit{sketch}$.
        \item For every $\sigma \in \{0,1\}^s$, let $X_{\sigma}$ denote the set of points $x \in X$ such that for $\tilde{x} = x - m$, $\Alg(\sigma, \tilde{x}) = \close$. For every non-empty $X_{\sigma}$, execute $\textsc{Core-Preprocess}(X_{\sigma}, k-1)$ and store the data structure node in $v.\textit{child}(\sigma)$.
    \end{enumerate}
\end{itemize}

\end{framed}
\caption{\textsc{Core-Preprocess} Subroutine.} \label{fig:preprocess}
\end{figure}

\begin{figure}[H]
\begin{framed}

\textsc{Core-Query}$(q, v)$.\\

\textbf{Input}: A query vector $q \in [\Delta]^d$, and a data structure node $v$ generated by the \textsc{Core-Preprocess} subroutine. Similarly to before, the scale $r > 0$, approximation $c > 1$, and parameter $\eps > 0$ is fixed. \\
\textbf{Output}: A point $x$ (coming from the preprocessed dataset), or ``fail''.

\begin{itemize}
    \item If $v$ is a leaf node, scan $v.\textit{data}$ and return the first point $x$ in $v.\textit{data}$ whose distance to $q$ is at most $cr$. Output ``fail'' if there are no such points.
    \item Otherwise, $v$ is not a leaf node. Let $m$ be the point stored in $v.\textit{med}$. If $\|q - m\|_p \leq (c-1)r/2$ and $x$ is stored in $v.\textit{close}$, return $x$ (since it must be within distance $cr$ of $q$). If $\|q-m\|_p \leq (c-1)r/2$ but $v.\textit{close}$ does not contain any point, output ``fail'' (since no $x \in X$ satisfies $\|q-x\|_p \leq r$).
    \item Otherwise, $\|q - m\|_p \geq (c-1)r/2$, and $v.\textit{sketch}$ contains the randomness for the sketch $(\sk, \Alg)$. Let $\bsigma = \sk(q-m)$, and output \textsc{Core-Query}$(q, v.\textit{child}(\bsigma))$.
\end{itemize}

\end{framed}
\caption{\textsc{Core-Query} Subroutine.} \label{fig:query}
\end{figure}

 In the subsequent claims, we let $\sfI(n)$ denote the initialization time of $(\sk, \Alg)$ when initialized for distribution on at most $n$ points in $\{-\Delta, \dots, \Delta\}^d$.
 Note that the necessary components are (i) computing the coordinate-wise median (which takes $O(nd)$ time) and (ii) generating and storing the randomness of $(\sk,\Alg)$ which takes $O(d)$ time.
 Furthermore, let $\sfT$ be the running time of applying the function $\sk$ and executing the algorithm $\Alg$.
 These steps take time $\tilde{O}(d) + 2^{O(p/c)} \cdot \log(1/\eps)$.
 Moreover, our sketches always have size $O(d)$, so we may bound this runtime by $\ot(d\log(1/\varepsilon))$.
 
\begin{claim}[Running Time of \textsc{Core-Preprocess}]
     For a dataset $X \subset [\Delta]^d$ of size $n$ and any $k \in \N$, the subroutine $\textsc{Core-Preprocess}(X, k)$ runs in time $2^{s \cdot k} \cdot O(nd + \sfI(n) + n\sfT)$. %
\end{claim}

The above claim follows by inspection of Figure~\ref{fig:preprocess}.
The arity of the tree is $2^s$, and the height of the tree is $k$, which implies there are at most $2^{s(k+1)}$ nodes in the tree.
At each node, the algorithm does $O(nd + \sfI(n) + n\sfT)$ work as discussed above, which implies the total running time.
Moreover, once we set $k = O(\log n)$, the total preprocessing time and space complexity of the data structure will be dominated by \smash{$d \cdot n^{O((p/c)\log(1/\eps))}$}, since $s=O((p/c)\log(1/\varepsilon))$ by Corollary \ref{corr:one-way}. 

\begin{claim}[Approximation Guarantee]
    A call to \textsc{Core-Query}$(q,\bv)$ where $\bv$ was generated from \textsc{Core-Preprocess}$(X, k)$ either outputs ``fail'', or a point $x \in X$ which is within distance at most $cr$.
\end{claim}
Clearly, the algorithm is designed such that if it outputs a point $x \in X$, then it must be at most a distance of $cr$ from $q$. If the algorithm is not able to find such a point, it outputs ``fail''.

\begin{lemma}[Success Probability of \textsc{Core-Preprocess} and \textsc{Core-Query}]
    For any dataset $X \subset [\Delta]^d$ of $n$ points and any query $q \in [\Delta]^d$, if there exists $x^* \in X$ with $\|x^* - q \|_p \leq r$, then
    \[ \Prx\left[\textsc{Core-Query}(q, \bv) \text{ doesn't fail when }\bv \leftarrow \textsc{Core-Preprocess}(X, k) \right] \geq (1 - \eps)^{k}. \]
    where the randomness is over the construction of the data structure $\bv$.
\end{lemma}

\begin{proof}
    The proof is by induction on $k$. In base case of $k = 0$: $\textsc{Core-Preprocess}(X, k)$ outputs $\bv$ and all of $X$ is stored in $\bv.\textit{data}$. The subroutine \textsc{Core-Query}$(q, \bv)$ then scans $\bv.\textit{data}$ and must encounter some point $x$ within distance $cr$ since $x^*$ is in $\bv.\textit{data}$. In this case, the data structure succeeds in this case with probability $1$. We thus assume for induction that the claim holds for and $k - 1$, and prove it for $k$.

    For $k > 0$, the method $\textsc{Core-Preprocess}(X, k)$ generates a node $\bv$ which initializes a sketch $(\sk, \Alg)$ for $\tilde{\mu}$, the uniform distribution over the set $\{\tilde{x} = x - m\}_{x \in X}$ where $m$ is the coordinate-wise median vector of $X$.
    If $\|q-m\|_p \leq (c-1)r/2$, the data structure cannot output ``fail'', since $x^*$ must be within distance $(c+1)r/2$ from $m$, so some point $x$ would be stored in $\bv.\textit{close}$.
    Thus, consider the case $\|q-m\|_p \geq (c-1)r/2$, and let $\bsigma^* = \sk(q - m)$.
    By Corollary \ref{corr:one-way},  we have $\Alg(\bsigma^*,\tilde{x^*}) = \close$ with probability at least $1 - \eps$. Suppose this event occurs, and consider any fixed draw of $(\sk, \Alg)$ where $\Alg(\sk(\tilde{p}), \sk(q-m)) = \close$.
    We let $X' = \left\{ x \in X : \Alg(\sk(q-m), \tilde{x}) = \close \right\}$, which satisfies $x^* \in X'$, and we consider the fixed value of $\sigma^* = \sk(q-m)$.
    When we execute \textsc{Core-Query}$(q, \bv)$, the third bullet is executed (since $\|q-m\|_p \geq (c-1)r/2$), and the data structure does not output fail whenever $\textsc{Core-Query}(q, \bv.\textit{child}(\sigma^*))$ does not output fail. Since $x^* \in X'$ the inductive hypothesis implies that, since $\bv.\textit{child}(\sigma^*) \leftarrow \textsc{Core-Preprocess}(X', k-1)$, $\textsc{Core-Query}(q, \bv.\textit{child}(\sigma^*))$ does not output fail with probability at least $(1-\eps)^{k-1}$. Therefore, the probability that both events occur, and that $\textsc{Core-Query}(q, \bv)$ does not output fail is at least $(1-\eps)^k$.
    \end{proof}

\begin{lemma}[Running Time of \textsc{Core-Query}]
    For any dataset $X \subset [\Delta]^d$ and query $q \in [\Delta]^d$, the expected running time of $\textsc{Core-Query}(q, \bv)$ where $\bv \leftarrow \textsc{Core-Preprocess}(X,k)$ is at most
    \[ O\left( k \cdot \sfT + d \cdot |X| \cdot (3/4)^{k} + d (k+1) \right).\]
\end{lemma}

\begin{proof}
    We proceed by induction on $k$. The base case occurs when $k = 0$, in which case $\bv.\textit{data}$ holds on to all of $P$ and $\textsc{Core-Query}(q, \bv)$ will scan $\bv.\textit{data}$ until it finds the first point within distance $cr$. Note that this takes time $O(d |X|)$. Assume for inductive hypothesis that the above claim holds for $k-1$, and we prove it for $k$.

    For $k > 0$, an execution of $\textsc{Core-Query}(q, \bv)$ for $\bv \leftarrow \textsc{Core-Preprocess}(X, k)$ will (i) take $O(d)$ time to compute the distance to $\bv.\textit{close}$, (ii) $O(d) + \sfT$ time to compute $\bsigma^* = \sk(q-m)$ for $m$ in $v.\textit{med}$, and then (iii) the time required to execute $\textsc{Core-Query}(q, \bv.\textit{child}(\bsigma^*))$, where $\bv.\textit{child}(\bsigma^*)$ is generated from a call to $\textsc{Core-Preprocess}(\bX', k-1)$, where $\bX'$ consists of all points $x \in X$ with $\Alg(\bsigma^*, \tilde{x}) = \close$.
    In the case that $\|q-m\|_p \leq (c-1)r/2$, the data structure outputs in $O(d)$ time. Otherwise, we have $\|q-m\|_p \geq (c-1)r/2$ and the we execute $\textsc{Core-Query}(q, \bv.\textit{child}(\bsigma^*))$ where $\bv.\textit{child}(\bsigma^*)$ was generated by $\textsc{Core-Preprocess}(\bX', k-1)$.
    
    By the inductive hypothesis and linearity of expectation, the total expected running time is %
    \[ O(d + \sfT) + O\left((k-1) \cdot (\sfT + d) + d \cdot \Ex_{(\sk, \Alg)}\left[ |\bX'| \right] \cdot (3/4)^{k-1} \right). \]
    Finally, notice that when $\|q-m\|_p \geq (c-1)r/2$, Corollary \ref{corr:one-way} implies
    \[ \Ex_{(\sk, \Alg)}\left[|\bX'| \right] = |X| \cdot \Prx_{\substack{(\sk, \Alg) \\ \tilde{\bx} \sim \tilde{\mu}}}\left[ \Alg(\sk(q-m), \sk(\tilde{\bx})) = \close \right] \leq |X| \cdot (3/4),\]
    which gives the desired expected running time. 
\end{proof}

\section{Lower Bound in the Certificate Model}\label{sec:lb}

In the proof of \cref{thm:ell-p-sketch}, we reduced the average-distortion sketch to the single-scale decision problem at a certain threshold $r$ (see \cref{lem:boost-prob-of-non-expansion}).
For the decision problem, our sketch proceeds by attempting to find a certain ``probabilistic certificates'' that the $\ell_p$-distance is strictly greater than $r$.
In the spirit of the ``probabilistic certificates'', in this section, we define a notion of certificate which guarantees that a pair of points is far (with high probability), and we will lower bound the space complexity of sketches which can find these certificates. 

\paragraph{Limitations of the Certificate Model} As this model suggests, what we will show is not a strict lower bound for the single-scale decision version of the average-distortion sketch.
Instead, we will prove a lower bound for a new (harder) problem defined in \cref{def:ell-infinity-decision-problem}. Compared to the single-scale decision problem, the differences or limitations are 1) that the new problem further requires the sketch to output a ``probabilistic certificate'' when it claims the vectors are far, 2) and the new problem is defined for a specific distribution (see Definitions~\ref{def:dist-lb} below).

We will show in \cref{cl:single-scale-to-certifier} that our sketch described in \cref{fig:sketch}, \cref{sec:ads-sketch} (with a minor modification and of size $\poly(p) \cdot 2^{p/c}$ after the modification) can solve the new problem. However, any sketch that solves it is of size at least $2^{\Omega(p/c)}$ (\cref{thm:lb}).
This provides evidence that one should avoid trying to find the ``probabilistic certificates'' when attempting to improve on the $2^{\Theta(p/c)}$ factor in the size of our sketches, {and further may give some reason to believe that a lower bound of $2^{\Omega(p/c)}$ space exists for the single-scale problem}.

We now give the distribution of the lower bound, before formally describing the sketching problem we prove requires $2^{\Omega(p/c)}$ space.

\begin{definition}[The distributions for the lower bound.]
\label{def:dist-lb}
 For any integer $p \in [2,\infty)$ and sufficiently large $c > 1$, generate $\bx^{(0)}, \bx^{(1)}, \dots, \bx^{(c-1)} \in \{0,1\}^{2^{p-1}}$ with the procedure defined as follows:
\begin{itemize}
    \item Sample a subset $\bJ_0 \subset [2^{p-1}]$ of indices of size $2^{p-2}$ uniformly at random. Set $\bx^{(0)}_{i} = 1$ for all $i \in \bJ_0$, and set $\bx^{(0)}_{i} = 0$ for all $i \not \in \bJ_0$.
    \item For $i \in \{1,2,\dots, c-1\}$, iteratively define $\bJ_i$ as follows.
    Sample a subset $\bJ_i$ of $\bJ_{i-1}$ uniformly at random of size $2^{-(p-2)/(c-1)} \cdot |\bJ_{i-1}|$\footnote{We assume $(p-2)/(c-1)$ is an integer, which is without loss of generality up to constant factors.}.\ \label[condition]{item:smaller-sets}
    Then, set $\bx^{(i)}_j = 1$ for all $\bj \in \bJ_i$, and set $\bx^{(i)}_{j} = 0$ for all $j \not \in \bJ_i$.
\end{itemize}

Let $\zeta$ be the distribution of $\left(\bx^{(0)}, \bx^{(1)}, \dots, \bx^{(c-1)}\right)$.
Let $\mu$ be the distribution of $\sum_{j} \bx^{(j)}$ where $\left(\bx^{(j)}\right)_j \sim \zeta$.
\end{definition}

\begin{remark}
    Throughout this section, we fix $p,c$ and use $\mu,\zeta$ as defined in Definition \ref{def:dist-lb}.
\end{remark}

For vectors $\bX,\bY$ drawn independently from $\mu$, notice that
$\|\bX - \bY\|_p \geq c$ with probability at least $1/2$.
By the construction of $\mu$, there exists an index $i \in [2^{p-1}]$ such that $\bX_i = c$.
Moreover, $\Pr(\bY_i = 0) = 1/2$.
Therefore, we must have $\Pr(\|\bX-\bY\|_p\geq c)\geq \Pr(\exists i\in[2^{p-1}],|\bX_i-\bY_i|\geq c) \geq 1/2$. 

\paragraph{Certificates.} Note that $\mu$, the distribution defined above, is supported on vectors in $\{0, \dotsc, c\}^{2^{p-1}}$.
Given $\bX, \bY \sim \mu$, we are interested in \emph{certificates} that prove the $\ell_p$ distance between $\bX$ and $\bY$ is greater than a positive constant $r$. In this section, we let $r \in (2^{1-1/p}, 2)$ be a universal parameter.
Notice that $\|\bX - \bY\|_p \geq cr/2$ with probability at least $1/2$ since $r < 2$.

\begin{definition}\label{def:cert}
    Given two vectors $x, y \in \{0, \dots, c\}^{2^{p-1}}$ and any $r < 2$, we say that the pair $(i, \ell) \in [2^{p-1}] \times [c-1]$ certifies that $\|x-y\|_p > r$ 
    whenever $x_i < \ell < y_i$.
\end{definition}
This definition of certificates for the distribution $\mu$ is motivated by the following reasons:
\begin{itemize}
    \item For any two vectors $x,y$ in the support of $\mu$ at a distance at least $r > 2^{1-1/p}$, there must exist a pair $(i, \ell)$ which certifies $\|x-y\|_p > r$.
    Recall that $\mu$ is supported on $2^{p-1}$ integer coordinates.
    If no pair $(i, \ell)$ exists as a certificate, then every coordinate differs by at most $1$, and hence $\|x-y\|_p \leq 2^{1 - 1/p} < r$.
    \item Suppose vectors $x,y$ are in the support of $\mu$, and a pair $(i, \ell)$ certifies that $\|x-y\|_p > r$ where $r < 2$ according to \cref{def:cert}.
    Then we must have $\|x-y\|_p \geq |x_i - y_i| \geq 2 > r$, and so the existence of such a pair actually does imply $\|x-y\|_p > r$.

    (The above two bullet points combined are saying, in the support of distribution $\mu$, $\ell_p$ behaves like $\ell_\infty$, in the sense that $\|x - y\|_p > r$ if and only if there exists a coordinate $i$ such that $|x_i - y_i| \geq 2$.)
    \item Finally, the average-distortion sketch described in \cref{lem:boost-prob-of-non-expansion} can output (with minor adaptation) a pair $(\bi, \bell)$ which certifies $\|\bX - \bY\|_p > r$ for $\bX,\bY \sim \mu$.
    We expand on this in \cref{cl:single-scale-to-certifier}.
\end{itemize}

\begin{claim}\label{cl:single-scale-to-certifier}
    Let $\calD$ denote the distribution over $(\sk, \Alg)$ given from \cref{lem:single-scale}. With a simplification to the algorithm $\Alg$ (tailored to $\mu$), the following is true:
    \begin{itemize}
        \item With probability $\Omega(1)$ over $\bX,\bY \sim \mu$ and $(\sk, \Alg) \sim \calD$, $\Alg(\sk(\bX),\sk(\bY))$ outputs a pair $(\bi, \bell)$. %
        \item For any $x, y \in \{0, \dotsc, c\}^{2^{p-1}}$, the probability over $(\sk, \Alg) \sim \calD$ that the algorithm outputs a pair $(\bi, \bell)$, but $y_{\bi} < \bell < x_{\bi}$ does not hold, is at most $ e^{-2^{\Theta(p/c)}}$.
    \end{itemize}
\end{claim}

\begin{proof}
We recall a few aspects of \cref{lem:single-scale} which we may tailor to the distribution $\mu$.
First, note that the distribution has median 0, so one does not need to recenter it. As the number of coordinates is $2^{p-1}$, we do not need to hash the coordinates and may store their identities directly (using $O(p)$ bits for each coordinate).
For any $x$ in the support of $\mu$, $\|x\|_p$ is always $c$, so we denote $\nu = \|x\|_p = c$.
We only consider $\sigma = 1$ since all values in the support are non-negative, and the embedding (dividing by $\bu_{\bi} \sim \Exp(1)$) perserves the signs. 

Let $(\sk,\Alg)$ be a sketching algorithm drawn from $\calD$. Since the norm of the vectors in the support of $\mu$ is fixed to $c$, \cref{step:1} of the output algorithm $\Alg$ in \cref{fig:sketch} never outputs $\far$ and always set $\gamma = 0$. Now, recall \cref{step:2} of the output algorithm $\Alg$ in \cref{fig:sketch}. Upon receiving the input $\sk(x),\sk(y)$, $\Alg$ outputs $\far$ if and only if 
\begin{enumerate}
    \item $|G_{x, \bu}(\tau(\nu, j-2), 1)| \leq (\sfk / 4) \cdot |G_{x,\bu}(\tau(\nu, j), 1)|$, equivalently, $s_{x,1} = 1$ (see Step~\ref{skstep:4} of Sketching Map $\sk$ for definition of $s_{x,1}$); and \label[condition]{eq:original-size-constraint}
    \item The first recorded coordinate $i \notin \bH_{y, \bu}(\tau(\nu, \bj-1), 1)$,  \label[condition]{eq:original-idx-out-of-range}
\end{enumerate}
One can easily verify the first condition always holds in the support of $\mu$. We modify $\Alg$, and define the new algorithm $\AlgCert$ which outputs a certificate.
After checking both the above conditions, $\AlgCert$ checks if the exponential random variable 
\begin{equation}
\bu_{\bi} > \frac{1}{2^{p+2}} \enspace. \label{eq:modified-third-check}
\end{equation}
If all three conditions are met, $\AlgCert$ outputs a certificate pair $(\bi, \bell)$ where $\bell = \floor{t\cdot \bu_{\bi}^{1/p} -  (r'/2) \cdot \left({\bu_{\bi}}/{\delta_1}\right)^{1/p}}$ for threshold $t = \tau(\nu, \bj)$. Otherwise, it outputs $\bot$.

Finally, we execute $\AlgCert$ for distance threshold $r' = 16 r$.
Consequently, the corresponding lower bound is for approximation $c / 16r$.

Now, we prove the first item of the claim. Recall that Lemma~\ref{lem:single-scale} shows that for $\|x\|_p=c$, $\by\sim\mu$, and $(\sk,\Alg)\sim\calD$, the sketching algorithm $\Alg(\sk(x),\sk(\by))$ outputs $\far$ with probability at least 1/4.
Since for all $x\in \supp(\mu)$, $\|x\|_p=c$, the statement further holds for any $\bx\sim \mu$.
Applying a union bound, our modified algorithm $\AlgCert$ outputs a certificate pair with probability at least
\begin{align*}
    &1-\Prx_{\substack{\bX,\bY \sim \mu \\ (\sk, \Alg) \sim \calD}}[\AlgCert(\sk(\bX),\sk(\bY))=\bot]\\
    \geq &1-\Prx_{\substack{\bX,\bY \sim \mu \\ (\sk, \Alg) \sim \calD}}\left[\Alg(\sk(\bX),\sk(\bY))=\close\vee \bu_{\bi}\leq\frac{1}{2^{p+2}}\text{ for the first coordinate }\bi\in\bH_{\bX,\bu}(\tau,1)\text{ for some }\tau\right]\\
    \geq & 1 - \Prx_{\substack{\bX,\bY \sim \mu \\ (\sk, \Alg) \sim \calD}}\left[\Alg(\sk(\bX),\sk(\bY))=\close\right]-\Prx_{\bu\sim(\Exp(1))^{2^{p-1}}}\left[\exists i\in[2^{p-1}],\bu_i\leq\frac{1}{2^{p+2}}\right]\\
    \geq& \frac{1}{4}-2^{p-1}\cdot\Prx_{\bu_1\sim\Exp(1)}\left[\bu_1\leq\frac{1}{2^{p+2}}\right] \\
    =&\frac{1}{4}-2^{p-1}\cdot(1-e^{-1/2^{p+2}}) \geq \frac{1}{4} - 2^{p-1}/2^{p+2} \geq \frac{1}{8} \enspace.
\end{align*}

We move on to prove the second item of the claim. Suppose upon receiving arbitrary vectors $x,y\in\{0,\ldots,c\}^{2^{p-1}}$, the algorithm $\AlgCert$ outputs a pair $(i,\ell)$. Thus, it must be the case that items \ref{eq:original-size-constraint}, \ref{eq:original-idx-out-of-range}, and condition (\ref{eq:modified-third-check}) from above are met. In particular, item \ref{eq:original-idx-out-of-range} indicates that $\AlgCert$ finds the threshold $t$ such that coordinate $i$ lies in $G_{x,\bu}(t,1)$ but not in $G_{y,\bu}(t-r'/(\delta_1/6)^{1/p},1)$. It asserts
\begin{align}
\frac{y_{i}}{\bu_{i}^{1/p}} + \frac{r'}{(\delta_1/6)^{1/p}} \leq t \leq \frac{x_{i}}{\bu_{i}^{1/p}} \enspace.\label{eq:assertions} 
\end{align}

The above assertion is incorrect with probability at most $e^{-\sfk/2}=e^{-2^{\Theta(p/c)}}$ (see proof of \cref{lem:close}); we will show that this is the only way the certificate can be incorrect.
On the other hand, by condition (\ref{eq:modified-third-check}), $\bu_i>2^{-(p+2)}$, which gives 
\begin{align}
\label{ineq:gap-larger-than-2}
\left({\bu_{\bi}}/{(\delta_1/6)}\right)^{1/p} \cdot r' \geq 2^{-(1+2/p)} \cdot r'  \geq r' / 8 = 2r > 2 \enspace.
\end{align}

Using (\ref{eq:assertions}) above, we get 
\begin{align*}
    y_i\leq t \cdot \bu_i^{1/p}-r'\cdot (\bu_i/(\delta_1/6))^{1/p} \enspace,
\end{align*}
which leads to $y_i<\ell$ immediately following inequality (\ref{ineq:gap-larger-than-2}) and the definition of $\ell=\floor{t\cdot \bu_{\bi}^{1/p} -  (r'/2) \cdot \left({\bu_{\bi}}/{(\delta_1/6)}\right)^{1/p}}$. 
Moreover, using the inequality (\ref{ineq:gap-larger-than-2}), we have 
\begin{align*}
    \ell = \left\lfloor t \cdot \bu_{i}^{1/p} -  \frac{r'}{2}\cdot \left({\bu_{i}}/{(\delta_1/6)}\right)^{1/p}\right\rfloor \leq  t \cdot \bu_{i}^{1/p} -  \frac{r'}{2}\cdot \left({\bu_{i}}/{(\delta_1/6)}\right)^{1/p}\leq t \cdot \bu_{i}^{1/p} -1 \enspace.
\end{align*}
Since $t \cdot \bu_{i}^{1/p} \leq x_{i}$, we obtain $\ell < x_i$ as desired.
\end{proof}

\cref{cl:single-scale-to-certifier} establishes that the single-scale sketch algorithm in \cref{fig:sketch}
 can be modified to output ``certificates of farness''.
 This motivates the following definition of a certification problem, and our goal for the remainder of this section is to lower bound the space complexity needed to solve the certification problem.

\begin{definition} 
\label{def:ell-infinity-decision-problem}The \emph{\textsc{Certification}} problem is defined as follows. The goal is to produce a distribution $\calD$ supported on tuples $(\sk, \Alg)$, where
\begin{itemize}
    \item $\sk \colon \{0, \dots, c\}^{2^{p-1}} \to \{0,1\}^s$, with $s$ being the space complexity of the sketch.
    \item $\Alg$ is an algorithm which takes two strings in $\{0,1\}^s$ as inputs and outputs $\bot$ or a certificate $(\bi,\bell)$ where $\bi \in [2^{p-1}], \bell \in [c-1]$.
\end{itemize}
We say $\calD$ $(\alpha,\delta)$-succeeds for \textsc{Certification} if and only if
\begin{itemize}
\item A certificate is output with probability at least $\alpha$, so
\begin{displaymath}
	\Prx_{\substack{\bX, \bY \sim \mu \\ (\sk, \Alg) \sim \calD}}\left[ \Alg(\sk(\bX),\sk(\bY)) \neq \bot \right] \geq \alpha. 
\end{displaymath}
\item If a certificate is output, it is incorrect with probability at most $\delta$.
That is, for any two points $x, y$ in the support of $\mu$, it holds that 
\begin{displaymath}
    \Prx_{\substack{(\sk, \Alg) \sim \calD}}[\Alg(\sk(x), \sk(y)) =  (\bi, \bell) \land \neg (y_{\bi} < \bell < x_{\bi}) ] \leq \delta
\end{displaymath}
\end{itemize}
\end{definition}%

We show a lower bound on the space complexity of solving \textsc{Certification} as stated in the following theorem.
\begin{theorem}
\label{thm:lb}
    For any $\delta \leq 1/((c-1) 2^{p + 2})$,\footnote{
    Whenever $c = O(p/\log p)$, the probability of outputting an incorrect certificate in Claim \ref{cl:single-scale-to-certifier} is $e^{-2^{\Theta(p/c)}} = 2^{-\Theta(p)}$ (and thus meets this condition on $\delta$).
    By increasing the space of the sketch of Claim \ref{cl:single-scale-to-certifier} by at most a $\poly(p)$ factor, we may only consider $c = O(p/\log p)$.
    } a distribution $\calD$ supported on tuples $(\sk, \Alg)$ which $(1/8,\delta)$-succeeds for \textsc{Certification} has space complexity at least $2^{\Omega(p/c)}$. 
\end{theorem}

We reduce \textsc{Certification} to a problem we call \rmi, a communication problem over $d$-length strings in the one-way two-party model.
Informally, in \rmi, Alice receives as input a uniformly random binary vector $\bx \in \{0,1\}^d$ with exactly $kt$ many $1$s, and Bob receives as input a uniformly random subset of indices $\bI\subset [d]$ of size $t$. Alice sends a message to Bob, and Bob's goal is to output a coordinate $i\in \bI$ such that $\bx_i = 0$.
However, if Bob is unable to find such a coordinate, he is also allowed to ``abstain'' and output $\bot$.
There are then two metrics upon which we judge the success of a protocol.
First, the probability $\alpha$ that Bob outputs an index $i\in \bI$, rather than abstaining and outputting $\bot$, and second the probability $\delta$ that when Bob outputs an index $i\in \bI$, it is ``incorrect'' in the sense that actually $\bx_i = 1$.
A protocol with high $\alpha$ and low $\delta$ is thus ``successful''; we will show that for $\alpha = \Omega(1)$ and $\delta =O(1/d)$, the problem requires $\Omega(k)$ bits of communication (Lemma \ref{lem:multi-index-is-hard}).

We now formally define the \rmi~problem.
In order to simplify the notation, for a parameter $n \leq d$, let $S_d(n)$ denote the set of binary vectors $\{0,1\}^{d}$ which have exactly $n$ many $1$s. For $n \leq d$, let $\calI_d(n)$ denote subsets of $[d]$ of size exactly $n$.
When the dimension $d$ is clear from context, we drop the subscript and write $S(n), \calI(n)$ respectively.

\begin{definition} \label{def:random-multiple-index}
    Fix a string length $d\in \N^+$.
    For $k, t \in \N^+$ with $kt \leq d/2$, $\emph{\textsc{Random-Multi-Index}}(k,t)$ is the following public-coin\footnote{In the public-coin communication model, Alice and Bob both have access to unlimited shared random bits, which are independent of both their inputs.} one-way communication problem:
    \begin{itemize}[nolistsep]
        \item Alice receives a uniformly random input $\bx \sim S_d(kt)$, and sends a message $\bm \in \{0,1\}^s$ to Bob.
        \item Bob receives a uniformly random input $\bI \sim \calI_d(t)$ and Alice's message $\bm$, and either outputs $\bot$ or an index $\bi \in \bI$.
    \end{itemize}
    We say $s = |\bm|$ is the communication complexity of the protocol. Moreover, we say that a protocol $\Pi$ for \textsc{Random-Multi-Index}$(k, t)$ $(\alpha, \delta)$-succeeds if the following to guarantees hold:
    \begin{itemize}[nolistsep]
        \item Bob outputs an index with probability at least $\alpha$.
        That is, with probability at least $\alpha$ over $\bx \sim S_d(kt)$, $\bI \sim \calI_d(t)$, and the randomness of $\Pi$, Bob outputs an index $\bi \in \bI$ (and not $\bot$).
        \item %
        With probability at most $\delta$ over $\bx \sim S_d(kt)$, $\bI \sim \calI_d(t)$, and the randomness of $\Pi$, Bob outputs an $\bi \in \bI$ such that $\bx_{\bi} = 1$.
    \end{itemize}
\end{definition}

\begin{lemma}
    \label{lem:reduce-to-multi-index}
    Let $d=2^{p-1}$.
    If there exists a distribution $\calD$ supported on tuples $(\sk, \Alg)$ that $(1/8,\delta)$-succeeds for \textsc{Certification} with space complexity $s$, then there exist $k, t$ such that $k \geq (d/2)^{1/(c-1)}$ and $kt \leq d/2$ for some $\delta\leq 1/d$, and a protocol $\Pi$ of communication complexity $s$, that $\left(\alpha,  \delta \right)$-succeeds, for $\alpha = 1/(8c)$, for \emph{\textsc{Random-Multi-Index}}$(k,t)$ with strings of length $d$.
\end{lemma}

\begin{proof}
    For a distribution $\calD$ that has space complexity $s$ and $(1/8,\delta)$-succeeds for \textsc{Certification}, there must exist a value $\ell_0 \in [c-1]$ such that 
    \begin{align}
        \label{eq:ell0}
        \Prx_{\substack{\bX, \bY \sim \mu \\ (\sk, \Alg) \sim \calD}}\left[\exists i\in [d] \text{ such that }\Alg(\sk(\bX),\sk(\bY)) = (i,\ell_0) \right] \geq \frac{1}{8(c-1)} \enspace, 
    \end{align}
    since $\Alg(\sk(\bX),\sk(\bY))$ is some certificate in the form of $(i,\ell)$ with probability at least $1 / 8$ and $\ell$ takes on $c-1$ possible values.
    Consider the following protocol $\tilde{\Pi}$
    for \textsc{Random-Multi-Index}$(k_0 , t_0)$ where $k_0 \coloneqq (d/2)^{1/(c-1)}, t_0 \coloneqq (d/2)^{1- (\ell_0+1)/(c-1)}$.
    \begin{itemize}
        \item Alice and Bob draw $(\sk, \Alg) \sim \calD$ using public randomness.
        \item Alice then performs the following steps:
        \begin{itemize}
            \item Given input $\bx \sim S(k_0 t_0)$, Alice draws $\{\bx^{(0)},\dots, \bx^{(c-1)}\} \sim \zeta$ conditioned on $\bx^{(\ell_0 -1)}=\bx$ (where $\zeta$ is defined in Definition \ref{def:dist-lb}).
            \item Alice computes $\bX = \sum_{j = 0}^{c-1} \bx^{(j)}$, and sends $\sk(\bX)$ to Bob.
        \end{itemize}    
        \item Upon receiving Alice's message, Bob:
        \begin{itemize}
            \item Given input $\bI \sim \calI(t_0)$, Bob creates a vector $\by \in S(t_0)$ such that $\by_{j} = 1$ if $j \in \bI$, and $\by_{j} = 0$ otherwise.
            \item Bob draws $\{\by^{(0)},\dots, \by^{(c-1)}\} \sim \zeta$ conditioned on $\by^{(\ell_0)}=\by$. He also computes $\bY = \sum_{j=0}^{c-1} \by^{(j)}$.
            \item Bob outputs $\bi$ if $\Alg(\sk(\bX), \sk(\bY)) = (\bi, \ell_0)$ and $\bi\in\bI$, and outputs $\bot$ otherwise.
        \end{itemize}
    \end{itemize}
In order to show that $\tilde{\Pi}$ $(\alpha,\delta)$-succeeds for $\rmi(k,t)$ for $\alpha=1/(8c)$ and $\delta=1/d$, recall that we need to prove the following two claims:
\begin{itemize}[nolistsep]
    \item Bob outputs $\bot$ with probability at most $1 - \alpha$.
    \item Given that Bob outputs an index $\bi\in\bI$, the probability that $\bx_{\bi} = 1$ is at most $\delta$.
\end{itemize}

The first claim follows from the following argument.
Bob outputs $\bot$ in the above algorithm if at least one of the following events occurs:
\begin{itemize}[nolistsep]
    \item $\Alg(\sk(\bX),\sk(\bY))$ outputs $(\bi,\bell)$ such that $\bell \neq \ell_0$. By our choice of $\ell_0$, this happens with probability at most $1 - 1/(8(c-1))$.
    \item $\Alg(\sk(\bX),\sk(\bY))$ outputs $(\bi,\ell_0)$ but $\bi \notin \bI$.
    This can only occur when $\neg (\bX_{\bi} < \ell_0 < \bY_{\bi})$, which occurs with probability at most $\delta \leq 1/d$ since $\calD$ $(1/8,\delta)$-succeeds for \textsc{Certification}.
\end{itemize}
Therefore, by the union bound, the probability of Bob outputting $\bot$ is at most $1 - 1/(8(c-1)) + 1/d \leq 1 - 1/8c$ as desired.

The second claim follows from the following argument.
Given that Bob outputs an index $\bi$, by construction of $\tilde{\Pi}$, it must be the case that $\Alg(\sk(\bX),\sk(\bY)) = (\bi, \ell_0)$ and $\bi \in \bI$.
By the second requirement of the \textsc{Certification} problem, the probability that $\neg (\bX_{\bi} < \ell_0 < \bY_{\bi})$ occurs is at most $\delta$, which implies that the probability that $\bX_{\bi} \geq \ell_0$ occurs is at most $\delta$.
By the construction of $\bX$, $\bX_{\bi} \geq \ell_0$ if and only if $\bx_{\bi} = 1$, and therefore the probability that $\bx_{\bi} = 1$ is at most $\delta$ as desired.\qedhere

\qedhere

\end{proof}

\newcommand{\Dec}{\mathrm{Dec}}

Lastly, we show $\textsc{Random-Multi-Index}$ is hard in Lemma~\ref{lem:multi-index-is-hard}, which follows from the following fact:
\begin{restatable}{fact}{infofact}[Similar to Fact 8.1 in \cite{de2010time}]
\label{fact:info-theory-lb}
Fix a set $U$, and let $\calZ$ be the uniform distribution over $U$.
Suppose $\Enc: U \times \{0,1\}^R \to \{0,1\}^{*}$ and $\Dec: \{0,1\}^{*} \times \{0,1\}^R \to U$ are a pair of (randomized) encoding and decoding functions such that for all $u\in U$ and $r\in \{0,1\}^R$, $\Dec(\Enc(u,r),r)=u$.
Then,
\[ \Ex_{\substack{\bZ \sim \calZ \\ \boldr \sim \{0,1\}^R}}\left[ |\Enc(\bZ, \boldr)| \right] \geq \log_2(|U|) - 3. \]

\end{restatable}
For completeness, we give a proof to Fact \ref{fact:info-theory-lb} in Appendix \ref{app:app}.

\begin{lemma} 
\label{lem:multi-index-is-hard}
    Consider $\rmi(k,t)$ on strings of length $d$. For any $\delta < 1 / (\log_2(d) \cdot d)$ and $k, t \in \Z^+$ with $kt \leq d/2$, any protocol that $(\alpha, \delta)$-succeeds for \emph{\textsc{Random-Multi-Index}$(k,t)$} has communication complexity at least $\alpha^2k/100 - O(1)=\Omega(\alpha^2   k)$.
\end{lemma}
\begin{proof}
    We will show that we can use a protocol $\Pi$ that $(\alpha, \delta)$ succeeds for \textsc{Random-Multi-Index}$(k,t)$ to construct a protocol that communicates the entire string $x \in \{0,1\}^{d}$ with $kt$ many $1$'s in a two-party,%
    one-way, public-coin communication setting.
    Sending such a string has a natural lower bound of $\log_2 \binom{d}{kt} - O(1)$
    bits in expectation, due to \cref{fact:info-theory-lb}, since the strings are drawn from a uniform distribution over a support of size $\binom{d}{kt}$.
    First, recall the guarantees of $\Pi$ that $(\alpha, \delta)$-succeeds for \textsc{Random-Multi-Index}$(k, t)$:
    \begin{itemize}
        \item With probability at least $\alpha$ over the draw of $\bx \sim S(kt)$, $\bI \sim \calI(t)$, and the randomness in the protocol, Bob outputs some index $\bi \in \bI$.
        \item With probability at most $\delta$ over the draw of $\bx \sim S(kt)$, $\bI \sim \calI(t)$, and the randomness in the protocol, Bob outputs $\bi \in \bI$ but $\bx_{\bi} = 1$.
    \end{itemize}
    
    Suppose the protocol $\Pi$ for $\rmi(k,t)$ is $(\bAlg_{A},\bAlg_{B})$.
    That is, upon receiving an input $\bx$, Alice sends $\bAlg_{A}(\bx)$ to Bob; upon receiving input $\bI$ and Alice's message $\bAlg_A(\bx)$, Bob outputs $\bAlg_B(\bAlg_A(\bx),\bI)$. Now consider the following protocol for Bob to identify $0$'s in Alice's string:
     \begin{itemize} 
         \item Bob initiates an empty set of indices $\bB_0$. Then, he draws $\bI_1, \bI_2, \dots, \bI_{g} \sim \calI(t)$, for $g = \frac{ d}{10t}$, %
         using the public coins.
         Iteratively, Bob executes his algorithm to compute $\bAlg_B(\bAlg_A(\bx),\bI_j)$ for all $j\in[g]$.
         If $\bAlg_B(\bAlg_A(\bx),\bI_j)=\bot$, define $\bB_j=\bB_{j-1}$; otherwise if $\bAlg_B(\bAlg_A(\bx),\bI_j)=\bi$ for some index $\bi$, define $\bB_j=\bB_{j-1}\cup\{\bi\}$.
    \end{itemize}
    From now on in this proof, to simplify the notations, We use $\Ex$ and $\Prx$ without any subscript to denote all expectations and probabilities over the draw of $\bx \sim S(kt), \bI_1,\ldots, \bI_g \sim \calI(t)$, and random bits $\br$ used by $\Pi$. 
    For notational ease, we will sometimes write $\bI = \{\bI_1,\ldots, \bI_g\}$
    We claim that 
    \begin{enumerate}
    \item $\Ex\left[|\bB_g|\right] \geq \frac{d}{100t} \cdot \alpha^2$;
    \item Let $\bo_{\bx, \bI, \br}=|\{i:i\in \bB_g,\bx_i=1\}|$ be the number collected indices that correspond to an 1 in Alice's string $\bx$. It holds that 
    $\Ex[\bo_{\bx, \bI, \br}] \leq \frac{d}{10t} \cdot \delta$.
    \end{enumerate}
    
    For the first claim, 
    let $\bz_j$ be an indicator random variable (over the draw of $\bI_j \sim \calI(t)$) for the event $\be_1^{(j)}$ that $\bAlg_B(\bAlg_A(\bx),\bI_j)=\bi$ for some $\bi\in\bI_j$ and the event $\be_2^{(j)}$ that $\bi \not \in \bB_{j-1}$ both being true. Note that $\Pr[\neg \be_2^{(j)}]\leq\Pr[\exists k\in \bI_j,k\in \bB_{j-1}]\leq |\bI_j|\cdot\frac{|\bB_{j-1}|}{d}=\frac{t|\bB_{j-1}|}{d}$.
    Since for all $j \in [\alpha\cdot g / 10]$, $|\bB_{j-1}| \leq j \leq~ \alpha\cdot g / 10$, we further have
    \[\Ex[\bz_j] =
    \Prx[\bz_j = 1] \geq \Prx \left[\be_1^{(j)}\right] - \Prx \left[\neg \be_2^{(j)}\right] \geq \alpha - |\bB_{j-1}|\cdot \frac{t}{d} > 9 / 10 \cdot \alpha.\]
    That gives 
    \[\Ex\left[|\bB_g|\right] = \Ex\left[\sum_{i \in [g]} \bz_i\right] \geq \Ex\left[\sum_{i \in [\alpha\cdot g/10]} \bz_i\right] \geq \alpha\cdot g/10 \cdot 9 /10 \cdot \alpha > \frac{d}{100t} \cdot \alpha^2.\]
    For the second claim, let $\bz'_j$ be the indicator variable (over the draw $\bx\in S(kt), \bI_1,\ldots, \bI_g \in \calI(t)$ and randomness of $\Pi$) that in the $j$'th execution, $\bAlg_B$ outputs $\bi \in \bI_j$ but $\bx_{\bi} = 1$. Note that \smash{$ \Ex[\bz'_j] \leq \delta$ for all $j \in [g]$} by the second guarantee of $\Pi$. Then, we have
    \[\Ex\left[\bo_{\bx,\bI,\br} \right] = \Ex\left[\sum_{j \in [g]} \bz'_j\right]  \leq g \cdot \delta = \frac{d}{10t} \cdot \delta\]
    Notice that Bob draws $\bI_1, \bI_2, \dots, \bI_{g}$ using the public coins, so Alice knows the indices collected by Bob and among the indices which $0$'s are ``correctly'' or ``incorrectly'' identified.
    To communicate the entire string, in addition to sending $\bAlg_A(\bx)$ to Bob according to $\Pi$, Alice also sends the following two kinds of messages.
    \begin{itemize}
        \item The indices that Bob ``incorrectly'' identified as $0$'s, i.e. $\{i:i\in\bB_g,\bx_i=1\}$, along with the number of such indices. 
        This requires at most $ \bo_{\bx,\bI,\br} \cdot \log_2(d) + \bo_{\bx,\bI,\br} + 1$ bits\footnote{Alice sends the number of the indices in order for Bob to identify the bits that encode the indices in the message. The way for Alice to communicate the number of the indices $\bo_{\bx,\bI,\br}$, instead of sending the number using $\log(d)$ bits, is to send $\bo_{\bx,\bI,\br}$ $1$'s followed by a $0$, which costs $\bo_{\bx,\bI,\br} + 1$ bits.};%
        \item The substring of $\bx$ at the indices not in $\bB_g$, which requires at most $\log_2 \binom{d - |\bB_g|}{kt}$\footnote{$ \binom{d - |\bB_g|}{kt}$ is always well defined, since $1 \leq t \leq 2^{p-2}$ and $d - |\bB_g| - kt = d - \frac{d}{10t} - t \geq 0$.} bits since the substring has length at most $d - |\bB_g|$ and contains at most $kt$ many $1$'s.
    \end{itemize}
    The above gives a protocol to communicate the entire string, whose expected complexity 
    must be at least $\log_2 {\binom{d}{kt}} - O(1)$ by \cref{fact:info-theory-lb}. We therefore have 
    \begin{align*}
       \Ex \left[s + \bo_{\bx,\bI,\br} \cdot (\log(d)_2 + 1) + \log_2 \binom{d - |\bB_g|}{kt} \right] \geq \log_2 \binom{d}{kt} - O(1)
    \end{align*}
    where $s$ is the communication complexity of $\Pi$. Finally, we can lower bound the communication complexity $s$ of the protocol:  
    {
    \begin{align*}
        s & \geq  \log_2 \binom{d}{kt} -\Ex \left[ \bo_{\bx,\bI,\br} \cdot (\log_2(d) + 1) + \log_2 \binom{d - |\bB_g|}{kt} \right] - O(1) \\
        & \geq \Ex \left[ \log_2\left( \binom{d}{kt} \middle/ \binom{d - |\bB_g|}{kt} \right) \right]  - O(1) &  \left( \Ex [\bo_{\bx,\bI, \br}] \leq \frac{d}{10t} \cdot \delta \text{ and } \delta \leq \frac{1}{\log_2(d) \cdot d} \right) \\
        & =\Ex \left[ \log_2 \frac{d(d - 1)\dots (d - |\bB_g| + 1)}{(d - kt)(d - kt - 1)\dots (d - kt - |\bB_g| + 1)} \right] - O(1) \\
        & \geq\Ex \left[ \log_2 \left(\frac{d}{d - kt}\right)^{|\bB_g|} \right]  - O(1)  &\left(\frac{d}{d-kt} \leq \frac{d-x}{d-kt-x},\forall x \in [0, d - kt)\right)\\
        & = \Ex \left[ |\bB_g| \right] \cdot \log_2 \left(\frac{kt}{d - kt} + 1\right) - O(1) \\
        & \geq \Ex \left[ |\bB_g| \right] \cdot \frac{kt}{d - kt}  - O(1) \\
        & \geq \alpha^2 k /100  - O(1) & { \left( \Ex\left[|\bB_g|\right] > \frac{d}{100t} \cdot \alpha^2 \right)}
    \end{align*}}where the penultimate inequality follows from the fact that $kt/(d-kt) \in [0,1]$ and for all $x\in [0,1]$, $\log_2(x+1) \geq x$. \qedhere

\end{proof}

\begin{proof}[Proof of Theorem~\ref{thm:lb}]
\newcommand{\commcomplex}{o(2^{(p-2)/(c-1)}/c^2)}
    Assume for contradiction that there exists a distribution $\calD$ with space complexity $\commcomplex$ that $
    (1/8,\delta)$-succeeds for \textsc{Certification} for $\delta = {1}/{(8(c-1)\cdot (p-1) \cdot 2^{p-1})}$. %
    Then, by Lemma \ref{lem:reduce-to-multi-index}, there exists a protocol $\Pi$ of communication complexity $\commcomplex$ that $(\alpha, \delta)$-succeeds for \textsc{Random-Multi-Index}$(k, t)$ with string length $2^{p-1}$, for $\alpha = {1}/{(8c)}, k = 2^{\frac{p-2}{c-1}}$, and some $t$ such that $kt \leq 2^{p-2}$.
    But by Lemma~\ref{lem:multi-index-is-hard}, $\left(\alpha, \delta \right)$-succeeding for \textsc{Random-Multi-Index}$(k, t)$ on strings of length $2^{p-1}$ requires communication {$\Omega(\alpha^2 k) = \Omega(2^{(p-2)/(c-1)}/c^2)$},
    which is a contradiction.
\end{proof}

\bibliographystyle{alpha}
\bibliography{references, waingarten}

\appendix
\section{Proof of Fact~\ref{fact:info-theory-lb}}
\label{app:app}
We restate the fact again for readers' convenience.  
\infofact*
\begin{proof}
    Fix any $r\in \{0,1\}^R$, and for notational ease let $\Enc(u) = \Enc(u,r)$ and $\Dec(\sigma) = \Dec(\sigma, r)$.
    Since $\Dec(\Enc(u))=u$ for all $u\in U$, $\Enc$ must be injective.
    So for any integer $s$, as there are only $2^s$ strings of length $s$, there can only be $2^s$ distinct $u\in U$ such that $|\Enc(u)|=s$.
    Similarly, for any $q$, there can be at most $v_q = \sum_{s=1}^{q-1}2^s = 2^{q}-2$ strings $u\in U$ such that $|\Enc(u)| < q$.
    For any integer $q$ such that $|U| \geq v_q$ there are then at least $|U|-v_q$ strings with length at least $q$.

    Thus, for any $q$ such that $|U| \geq v_q$, we have
    \begin{align*}
        \Ex_{\bZ \sim \calZ}[|\Enc(\bZ)|] &\geq \frac{1}{|U|}\left( \sum_{s=1}^{q-1} s\cdot 2^s + q\cdot (|U|-v_q) \right) \\
        &= \frac{1}{|U|}\left( (q-1)2^q - 2^q + 2 + q\cdot (|U|-2^q+2) \right) \\
        &= q + \frac{1}{|U|} \left( 2 -2^{q+1} +2q \right) \\
        &\geq q - 2
    \end{align*}
    where the last inequality follows from the fact that $|U| \geq v_q = 2^q - 2 = -(2 - 2^q)$.
    For $q'=\floor*{\log_2(|U|)}$, $|U| \geq 2^{q'}-2 = v_{q'}$, so the above inequality holds for $q=q'$.
    Thus, as $q' \geq \log_2(|U|)-1$, we have
    \[\Ex_{\bZ \sim \calZ} \left[|\Enc(\bZ)|\right] \geq \log_2(|U|) - 3.\]
    Since $r\in \{0,1\}^{R}$ is an arbitrary string, the fact follows.
\end{proof}

\end{document}